\documentclass[pre,twocolumn,showpacs,floatfix,superscriptaddress,aps]{revtex4}
\usepackage{amsmath}
\usepackage{makeidx}
\usepackage{amssymb}
\usepackage{graphicx}

\usepackage[usenames]{color}
\usepackage[normalem]{ulem}

\newcommand{\beq}{\begin{equation}}
\newcommand{\eeq}{\end{equation}} 
\newcommand{\beqa}{\begin{eqnarray}}
\newcommand{\eeqa}{\end{eqnarray}} 
\begin{document}


\title{Elastic collision and molecule formation of  spatiotemporal light bullets  in a cubic-quintic nonlinear medium}

\author{ S. K. Adhikari\footnote{adhikari@ift.unesp.br; URL: http://www.ift.unesp.br/users/adhikari}
} 
\affiliation{
Instituto de F\'{\i}sica Te\'orica, UNESP - Universidade Estadual Paulista, 01.140-070 S\~ao Paulo, S\~ao Paulo, Brazil
} 

\begin{abstract}

We consider  the statics and dynamics  of a stable, mobile   three-dimensional (3D) spatiotemporal light bullet
in  a cubic-quintic nonlinear medium with a focusing cubic nonlinearity above a critical 
value and any defocusing quintic nonlinearity. 
The 3D light bullet  can  propagate with a constant velocity in any direction. 
 Stability of the light bullet under a small perturbation is established numerically.
We consider frontal collision between two light bullets with different relative velocities.
At large velocities the collision is elastic with the bullets emerge after collision with practically 
no distortion. At small velocities two bullets coalesce to form a bullet molecule. 
At a small range of intermediate  velocities  the localized bullets could form a 
single entity which expands indefinitely leading to a destruction of the bullets
after collision. 
The present study is based on an analytic Lagrange variational approximation and a full numerical solution of the 3D nonlinear 
Schr\"odinger equation.

\end{abstract}

\pacs{05.45.-a, 42.65.Tg, 42.81.Dp}

\maketitle

 \section{Introduction}
 
A bright soliton is a self-bound object that travels at a constant velocity  
in one dimension (1D), due to a cancellation of  nonlinear attraction and defocusing forces \cite{book,sol}.   The 1D soliton in a cubic Kerr medium  has been observed  in nonlinear optics \cite{book,sol} and in Bose-Einstein condensates \cite{rmp}. 
 Specifically, optical temporal \cite{temporal}  and spatial \cite{1996}
solitons were observed  for a cubic Kerr nonlinearity.
However, a three-dimensional (3D)   spatiotemporal soliton
cannot be formed in isolation with a cubic Kerr nonlinearity 
due to collapse \cite{collapse,book}. 
The same is true about  a two-dimensional (2D)  spatial soliton  with a Kerr nonlinearity.
However, the solitons  can be stabilized in higher dimensions
 for  a saturable  \cite{lbth,collapse}
 or a modified nonlinearity 
\cite{quartic}, or by a nonlinearity \cite{sadhan} or dispersion \cite{sadhan2} management  among other possibilities \cite{malomed,malomed2}. 
A 2D   spatiotemporal optical soliton has 
been observed \cite{stempo} in a saturable nonlinearity generated by the cascading of quadratic
nonlinear processes. The generation of a 2D spatial soliton in
an attractive cubic and repulsive quintic medium has been suggested \cite{2dcqth} and realized 
experimentally \cite{2dcqexp}. The generation of a stable 2D vortex soliton in a cubic-quintic medium has been suggested \cite{2dcqvor}.

Light bullets \cite{collapse}  are localized 3D
pulses of electromagnetic energy that can travel through a medium and retain their spatiotemporal shape due to  a balance between the non-linear self-focusing and spreading effects of the medium in which the pulse beam propagates. Such  light bullets are unstable and collapse in a cubic Kerr medium.
 Light bullets were realized experimentally in arrays of wave guides \cite{lbexp}. There were many theoretical $-$ numerical and analytical $-$  studies which established robustness and approximate solitonic nature of the light bullets  using the 3D nonlinear Schr\"odinger (NLS) equation \cite{book} with a modified nonlinearity \cite{quartic,lbth}, dissipation \cite{gl}, and/or dispersion \cite{sadhan2}. A saturable nonlinearity leads to stable optical bullets \cite{lbth}. Nonlinear dissipation in the complex cubic-quintic Ginzburg-Landau equation also stabilizes the bullets \cite{gl}.
Dispersion management can stabilize light bullets in a medium with cubic nonlinearity
 \cite{boris1}. There has been variational study of light bullets in a cubic-quintic medium \cite{cq} where a condition of stability was obtained. Another study suggested a way of 
the  stabilization of  light bullets in  a cubic-quintic medium
by a periodic variation of diffraction and dispersion \cite{stacq}.

In this paper we study the formation of a    3D spatiotemporal
light bullet \cite{lbth,collapse} in a cubic-quintic medium  for a defocusing quintic nonlinearity and a 
focusing cubic nonlinearity   as the ground state of  the 3D  NLS equation. 
A cubic-quintic medium is of experimental interest also. The study with a polydiacetylene paratoluene sulfonate  crystal in the wavelength region near 1600 nm shows that 
the refractive index versus input intensity correlation leads to a cubic-quintic form of 
nonlinearity in the NLS equation \cite{book,cqexpt}. The cubic-quintic nonlinearity also 
arises in a low intensity expansion of the saturable nonlinearity used in the pioneering 
study of light bullets \cite{lbth}. In this study of light bullets in a cubic-quintic medium 
we find that a stable light bullet can be formed for the cubic focusing nonlinearity above a critical value  in the 3D NLS equation for any finite quintic defocusing nonlinearity. 
The statical properties  of the light bullet is studied using a Lagrange variational analysis
and a complete numerical solution  of the   3D NLS equation.   The variational and numerical 
results are found to be in good agreement with each other. The stability of the light bullet is established numerically 
under a small perturbation introduced by changing the cubic nonlinearity by a small amount, while the bullet is found to execute sustained breathing oscillation.

The light bullet   can move freely without deformation along any direction  with a constant velocity. 
We also study the frontal collision between two light bullets. Only the collision between two integrable 1D solitons is truly elastic \cite{book}. As the dimensionality of the  
soliton is increased such collision is expected  to become inelastic with loss of energy in 2D and 3D.  
In the present numerical simulation of frontal collision between two light bullets
in different parameter domains of nonlinearities and velocities three distinct scenarios are found to take place. At sufficiently large velocities  the collision is found to be quasi elastic when the two bullets emerge after collision with practically no deformation. 
  At small   velocities 
 the collision   
is inelastic and  the bullets  form a single bound entity in an excited state
 and  last for ever and execute oscillation. We call this a bullet molecule. 
In a small domain of intermediate velocities, the bullets coalesce to form a single entity, which expands indefinitely  leading to the destruction of the bullets.

 .

We present the 3D NLS equation used in this study  in Sec. 
\ref{II}. 
In Sec. \ref{III} we present the numerical results for stationary profiles of  3D spatiotemporal light bullets.  We present numerical tests of stability of the light  bullet  under a small perturbation. 
The quasi-elastic nature of collision 
of two solitons at large velocities and bullet molecule formation at low velocities 
 are  demonstrated by real-time simulation. 
We end with a summary of our findings in Sec. \ref{IV}.

\section{Nonlinear Schr\"odinger equation: Variational formulation}
 
\label{II}

{
In nonlinear fibre optics a general 3D NLS equation  can usually be written as \cite{book,agarwal}
\begin{align}& \,
{\Big [}  i \frac{\partial }{\partial z} +\frac{1}{2 \beta_0}\nabla_\perp^2+\frac{\beta_2}{2}\frac{\partial^2}{\partial \mathrm{t}^2}+ \gamma \vert A \vert^2
{\Big ]}  A(\mathrm{x,y,t})=0,
\label{eq0}
\\
& \nabla_\perp ^2= \frac{\partial^2}{\partial \mathrm{x}^2}+
\frac{\partial^2}{\partial \mathrm{y}^2},
    \end{align}
where 
$\beta_2$ is dispersion parameter and can be positive or negative with magnitude of the order of { 10$^{-3}\sim 10^{-2}$ ps$^2$/m
\cite{agarwal}, } the nonlinear parameter 
$\gamma$  has unit { 
$ W^{-1}$m,} the unit of $|A|^2$ is {$W $m$^{-2}$, }
$\beta_0=2\pi n_0/\lambda$ is the propagation parameter \cite{book}, where $n_0$ is the refractive index and $\lambda$ is the wave length of the beam. The function $A$ describes the evolution of the beam envelop. For a spatiotemporal soliton it is useful to define the characteristic lengths for dispersion 
 $L_{DS}\equiv \tau^2/|\beta_2|$, and diffraction $L_{DF}\equiv 2n_0\pi \rho^2 /\lambda$ where $\rho$ is the radius of the beam, and $\tau$ is the 
time scale of the soliton \cite{malomed}. For an equilabrated propagation of the spatiotemporal soliton these two lengths are to be equal $-$  $L_{DS}=L_{DF}\equiv L_D$ $-$ yielding
$
\rho^2 = L_D \lambda/({2 \pi n_0}).
$
Now by scaling  we define the following dimensionless variables \cite{agarwal}
\begin{align}\;
 x=\frac{\mathrm{x}}{\rho},\;  y=\frac{\mathrm{y}}{\rho}, \; t=\frac{\mathrm{t}}{\tau},\; z =\frac{\mathrm{z}}{L_D},\;
{ \phi=\frac{A\rho}{\sqrt P_0}, \; p=\frac{\gamma P_0 L_D}{\rho^2}.}\label{unit}
\end{align}
The scale $P_0$ is chosen, so that $\int|\phi|^2 dx dy dt =1.$
} Using the dimensionless variables we obtain  
the following dimensionless NLS equations  with self-focusing cubic and self-defocusing 
quintic nonlinearity
    \cite{book}
\begin{align}& \,
{\Big [}  i \frac{\partial }{\partial z} +\frac{1}{2} \left( \frac{\partial^2}{\partial x^2} +\frac{\partial^2}{\partial y^2} 
 +\frac{\partial^2}{\partial t^2}\right)+ p \vert \phi \vert^2
- q \vert \phi \vert^4
{\Big ]}  \phi({\bf r},z)=0,
\label{eq1}
\end{align}
in scaled units 
where ${\bf r}\equiv \{x,y,t\}$,     $p$ is the  cubic and $q$ the quintic 
nonlinearity. Here $z$ is the propagation distance, $x,y$ denote transverse extensions, and 
$t$ the time. 
The plus sign before $|\phi|^2$ in Eq. (\ref{eq1})
denotes a self-focusing cubic nonlinearity. The quintic nonlinearity of strength $q$ with a negative sign 
denote self-defocusing.
 
{To have an idea of the length and time scales concerned, let us consider the case of an infrared beam of wave length 
$\lambda = 1$ $\mu$m, and take a nonlinear medium of { $\beta_2= 10^{-2}$ ps$^2$/m,} and a time scale $\tau=60$ fs.  Then the propagation length $L_D=36$ cm and the beam width $\rho \approx 239$ $\mu$m. These numbers are quite similar to those in an
experiment on spatiotemporal soliton in a planar glass waveguide \cite{sptex}.
 In this paper we quote the results in dimensionless units,
 which can be easily converted to actual experimental units following these guidelines.}
  
 For an analytic understanding we consider the Lagrange variational formulation of the 
formation of a light bullet. In this spherically symmetric problem, convenient analytic
Lagrangian variational approximation of Eq. (\ref{eq1}) can be obtained with
the following Gaussian ansatz for the wave function \cite{pg}
\begin{align}\label{eq2}
 \phi({\bf r},z)=\frac{\pi^{-3/4}}{w^{3/2}(z)}\exp\left[-\frac{r^2}{2w^2(z)}+i\alpha(z)r^2\right],
\end{align}
where $r^2=x^2+y^2+t^2$, $w(z)$ is the width and $\alpha(z)$ is the chirp.
 The Lagrangian density corresponding to Eq. (\ref{eq1}) is given by
\begin{align}\label{eq3}
{\cal L}({\bf r},z)= \frac{i}{2}\left[  \phi({\bf r},z)\frac{\partial \phi^*({\bf r},z)}{\partial z}- \phi^*({\bf r},z)\frac{\partial \phi({\bf r},z)}{\partial z}\right]\nonumber
\\+\frac{|\nabla \phi({\bf r},z) |^2}{2}-\frac{p}{2}| \phi({\bf r},z)|^4
+\frac{q}{3}| \phi({\bf r},z)|^6.
\end{align}
Consequently, the effective Lagrangian $L\equiv  \int {\cal L}({\bf r},z) d{\bf r}$ becomes
\begin{align}\label{eq4}
L=\frac{3}{2}w^2\dot \alpha+ \frac{3}{4w^2} +3 w^2\alpha^2-\frac{p\pi^{-3/2}}{4\sqrt 2  w^3}
+\frac{q\pi^{-3}}{9\sqrt 3  w^6},
\end{align}
where the overhead dot denotes the $z$ derivative. { The actual physical dimension J/m  of this dimensionless Lagrangian $L$ 
can be  restored upon multiplication by the factor $\tau P_0/L_D$.} 
The Euler-Lagrange equation for this Lagrangian yields the following ordinary differential 
equation for the width $w$:
\begin{align}\label{eq5}
\ddot w= \frac{1}{w^3} -\frac{p (2\pi)^{-3/2}}{w^4}+\frac{4q\pi^{-3}}{9\sqrt 3 w^7}.
\end{align}
The energy of the stationary bullet is the Lagrangian (\ref{eq4})  with $\alpha =0$, e. g.,
\begin{equation}
 E  =\frac{3}{4w^2}-\frac{p\pi^{-3/2}}{4\sqrt 2  w^3}+\frac{q\pi^{-3}}{9\sqrt 3  w^6}.\label{eq6a}
\end{equation}
The width $w$ of a stationary bullet is obtained by setting the right-hand-side of Eq. (\ref{eq5})   to zero:
\begin{align}\label{eq7}
 \frac{1}{w^3} -\frac{p (2\pi)^{-3/2}}{w^4}+\frac{4q\pi^{-3}}{9\sqrt 3 w^7}=0,
\end{align}
which is the condition for a minimum of energy $E$
$-$ $dE/dw=0,      d^2E/dw^2>0$    .
Without the quintic term ($q=0$) the bullet of width  $w=p/(2\pi)^{3/2}$ is tantamount to an unstable Towne's soliton \cite{towne}. 
  We will demonstrate that for stability  a 
non-zero quintic term ($q\ne 0$) is necessary.
For $q>0$, Eq. (\ref{eq7}) has solution for the cubic nonlinearity $p$ above a critical value $p_{\mathrm{crit}}$, which is the threshold for the formation of the bullet.

\begin{figure}[!t]

\begin{center}
\includegraphics[width=.8\linewidth,clip]{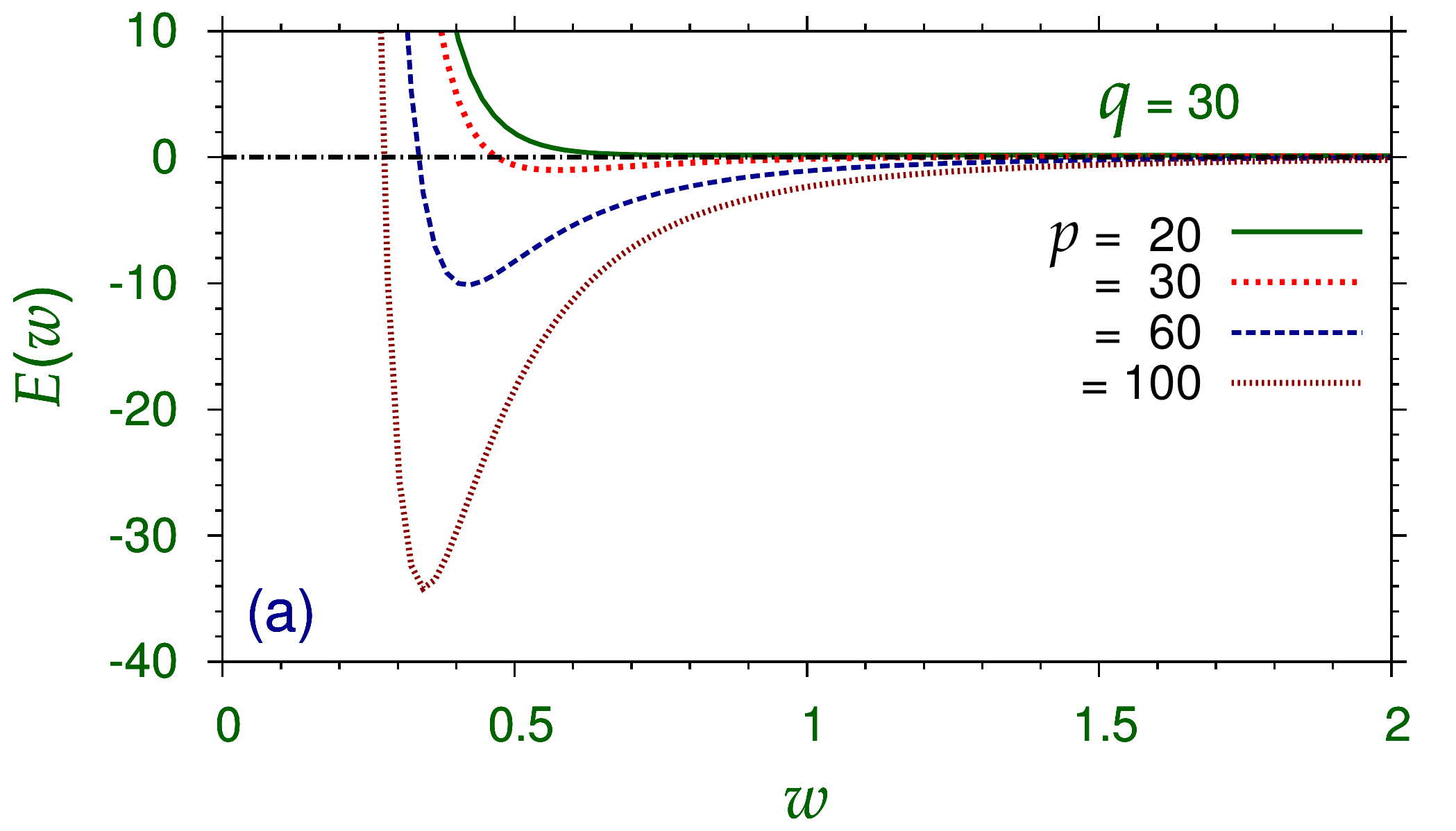} 
\includegraphics[width=.8\linewidth,clip]{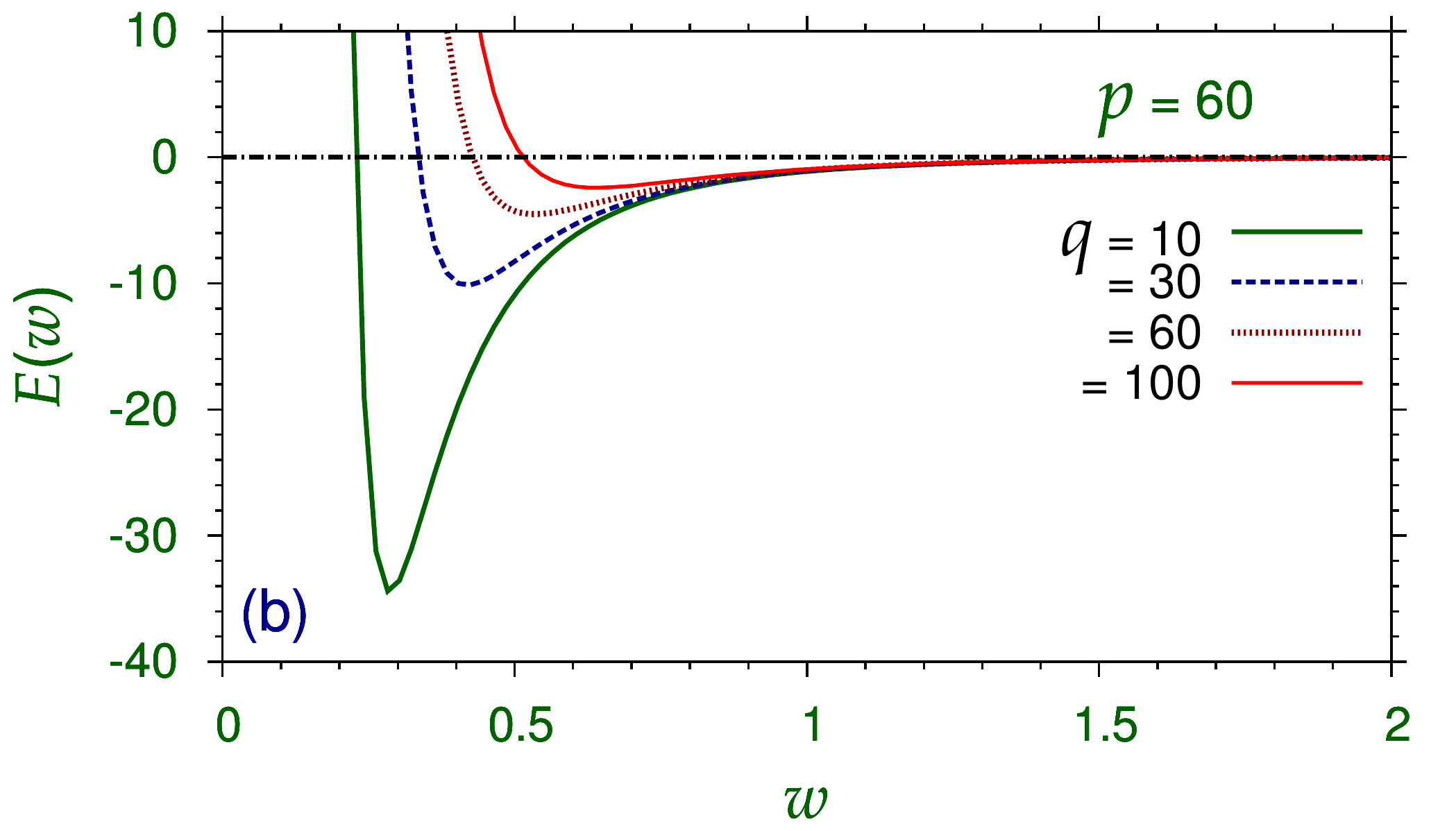}
\includegraphics[width=.8\linewidth,clip]{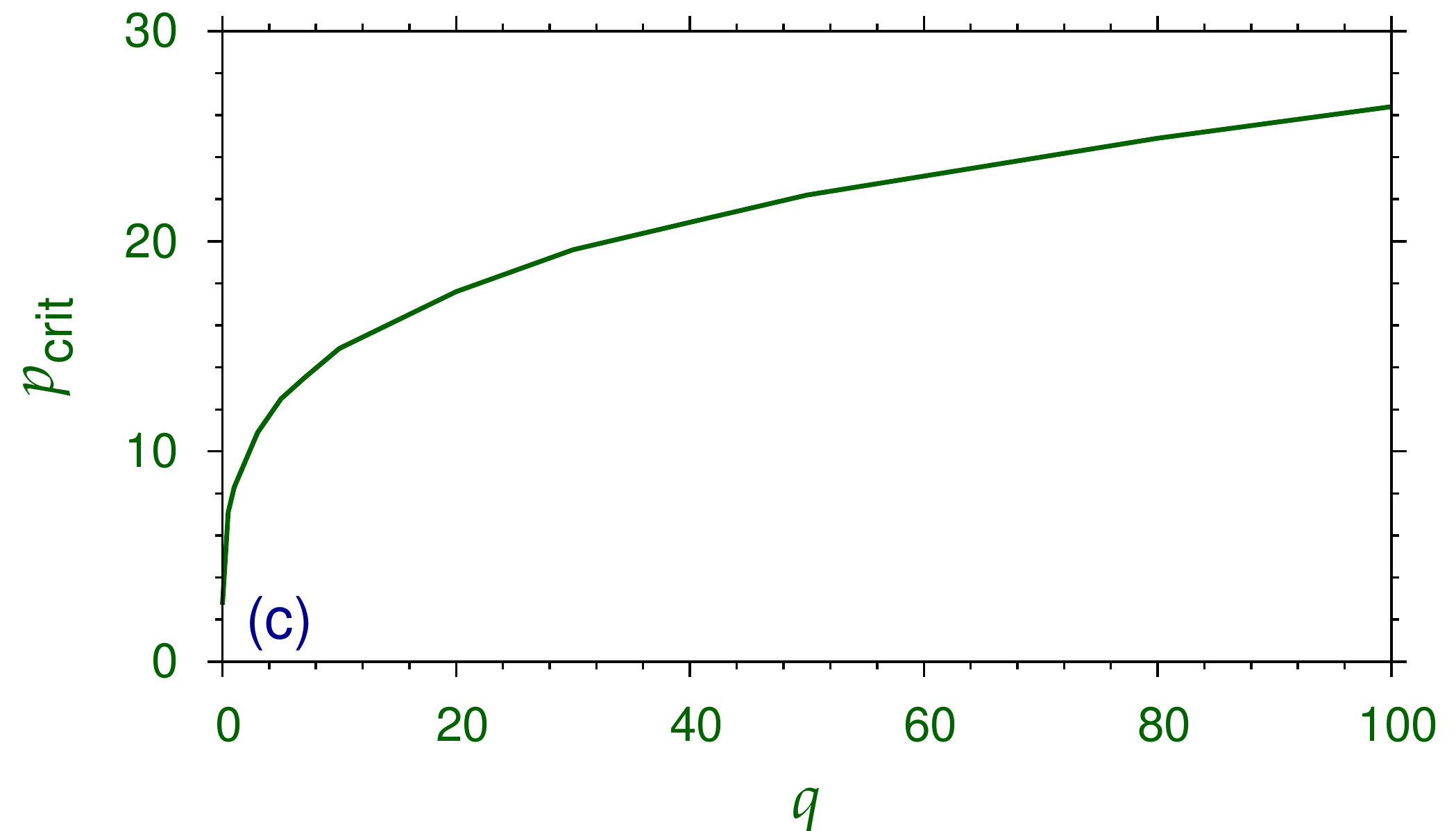} 
\includegraphics[width=.8\linewidth,clip]{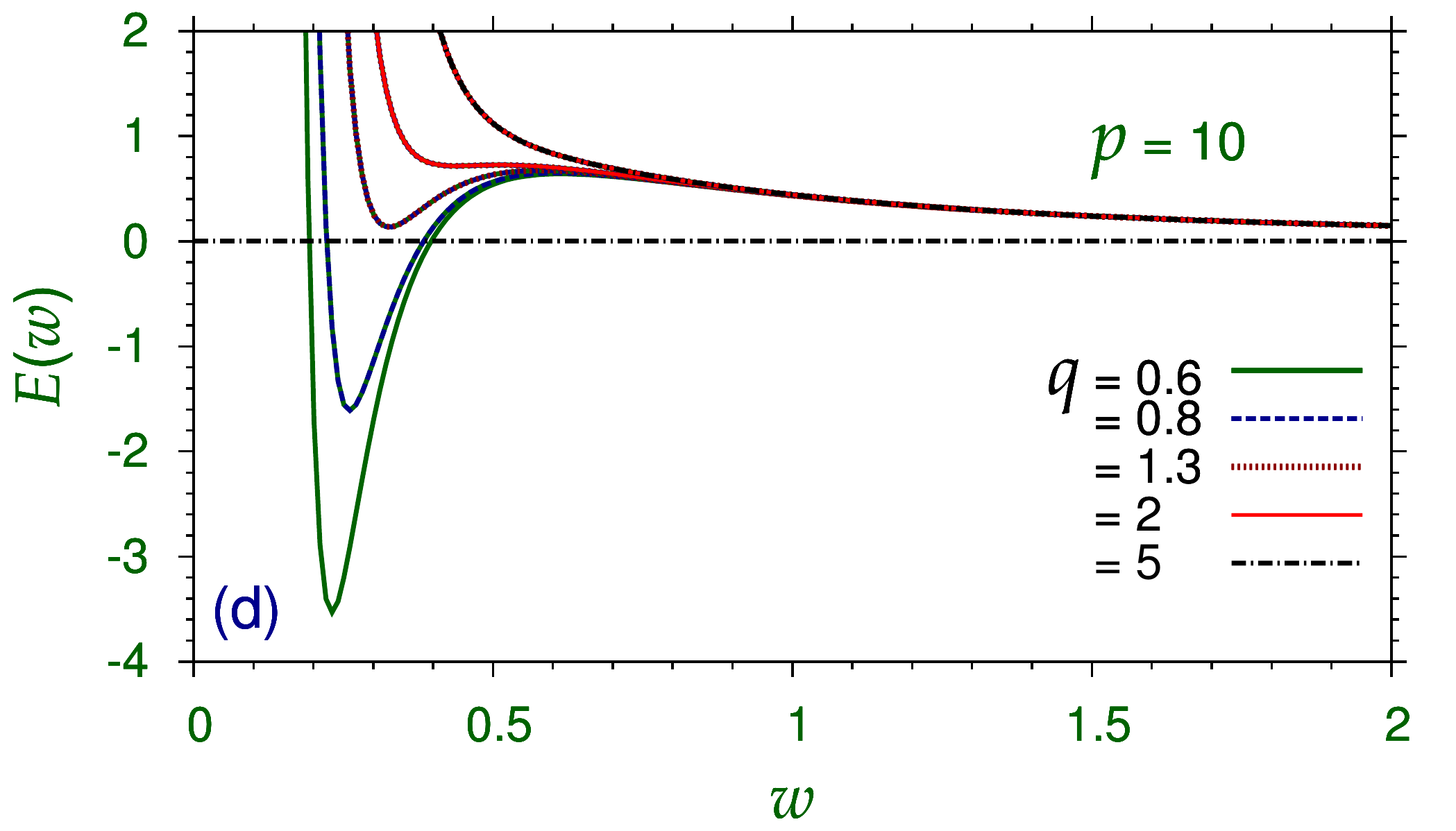} 
\caption{ (Color online) (a) Variational energy versus width ($E-w$) plot for different cubic nonlinearities 
$p (=20,30,60,100)$ and quintic nonlinearity $q=30.$
(b)The same for different quintic nonlinearities  
$q$ $ (=10,30,60,100)$ and cubic nonlinearity  $p=60.$ (c)  Variational critical cubic nonlinearity $p_{\mathrm{crit}}$ for light bullet formation, obtained from Eq. (\ref{eq7}),
for different values of quintic nonlinearity $q$. (d) Variational energy versus width ($E-w$) plot for different quintic nonlinearities 
$q(=0.6, 0.8, 1.3,2, 5)$ and cubic nonlinearity $p=10.$
}\label{fig1} \end{center}

\end{figure}

\section{Numerical Results}

\label{III}

Unlike in the 1D case, the 3D NLS equation (\ref{eq1})
 does not have analytic solution and different numerical methods, such as split-step Crank-Nicolson \cite{CPC} and Fourier spectral \cite{jpb}
methods, are used for its solution.
 Here we solve it numerically
by the split-step 
Crank-Nicolson method 
  in Cartesian coordinates  
using a  $\bf r$ $=\{x,y,t\}$  step of  $ 0.025$
and a $z$ step of  $ 0.0002$ \cite{CPC}. The number of ${\bf r}$ discretization points for each components is 256. 
There are different C and FORTRAN programs for solving the NLS-type equations \cite{CPC,CPC1}
and one should use the appropriate one. 
{We use both imaginary- and real-$z$ propagations \cite{CPC} in the numerical solution of the 3D NLS equation. The imaginary-$z$ propagation is appropriate  to find the stationary lowest-energy profile of the bullet. This method replaces $z$ by a new variable $\hat z \equiv i z$ and consequently Eq. (\ref{eq1}) becomes completely real 
and a $z$-iteration of this equation leads to the lowest-energy state with high accuracy. The real-$z$ propagation involve complex variable and hence is more complicated and less accurate. However, the real-$z$ propagation 
yields the 
propagation dynamics  of the bullet. In the imaginary-$z$ propagation, as  the propagation variable $z$ is replaced by the (unphysical) variable $\hat z$, this method cannot lead to the propagation dynamics of the bullet.   } In the imaginary-$z$ propagation the initial  state was taken as  in Eq. (\ref{eq2})
with $\alpha(z)=0$ and the width $w$ set equal to  the variational solution obtained by solving Eq. (\ref{eq7}). The convergence will be quick if the guess for the width $w$ is close to the final width. 
All stationary profiles of the bullets are calculated by imaginary-$z$ propagation. 
The dynamics and collision are then studied by real-$z$ propagation using the initial profile obtained in the  imaginary-$z$  propagation.

\begin{figure}[!t]

\begin{center}

\includegraphics[width=\linewidth,clip]{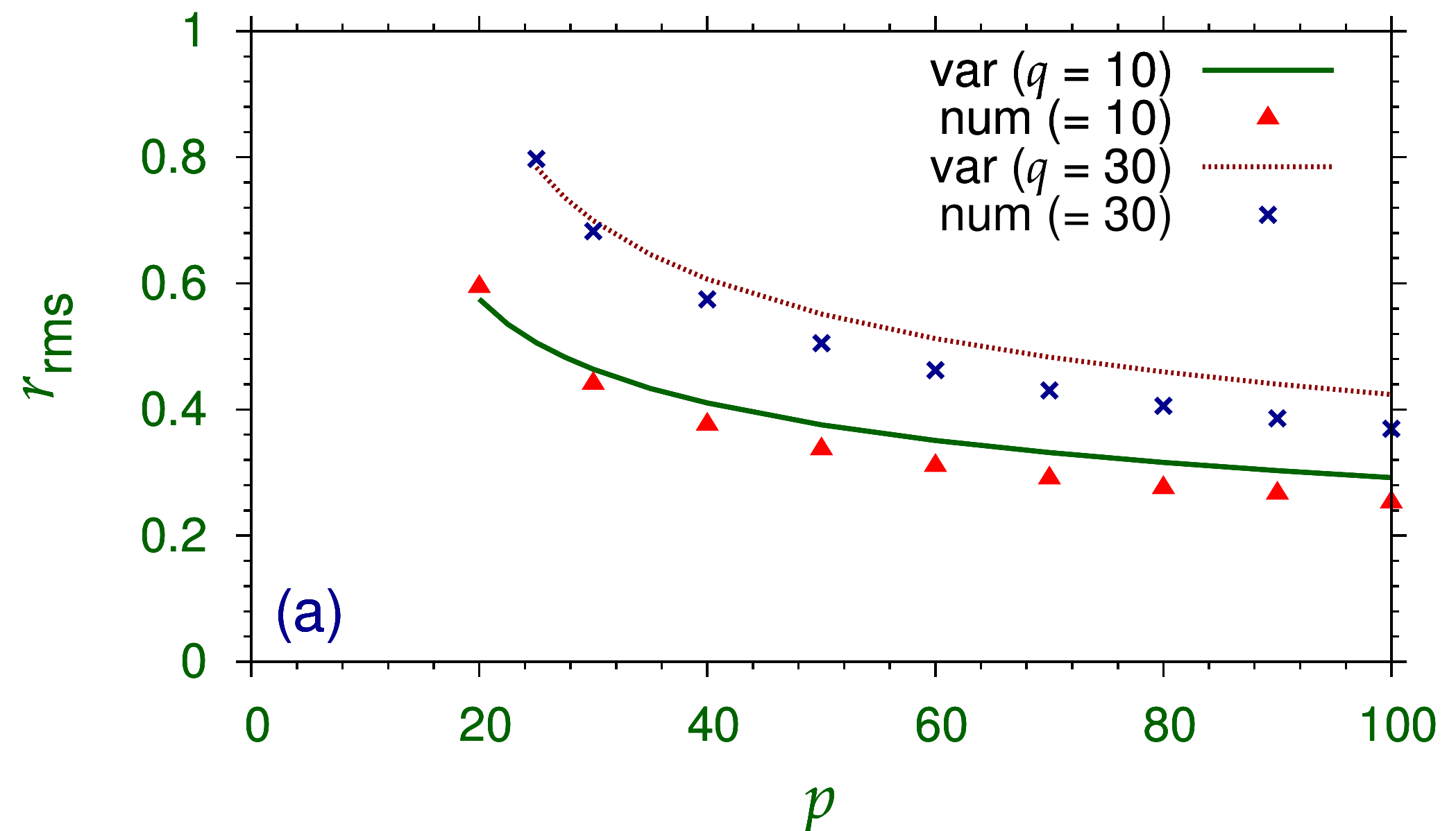}
\includegraphics[width=\linewidth,clip]{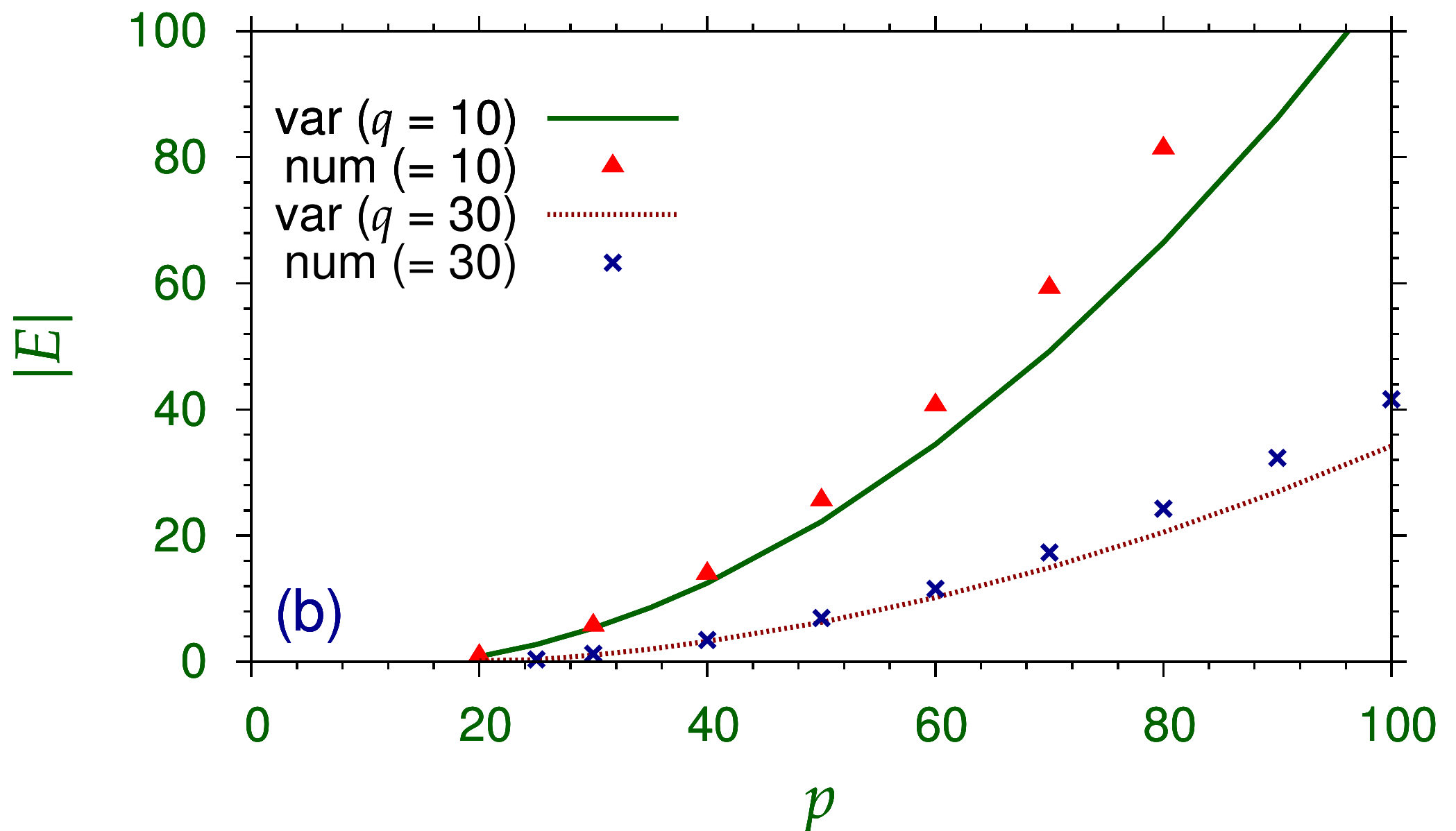}

\caption{ (Color online)
  Variational (var) and numerical (num) (a)  rms radius    and (b) energy $|E|$ versus  cubic nonlinearity $p$  of a light bullet 
for  two 
different quintic nonlinearities $q$ $ (=10,30)$.
}\label{fig2} \end{center}

\end{figure}

The stable bullet corresponds to an energy ($E$) minimum as given by  Eq. 
(\ref{eq7}).  In Figs. \ref{fig1}(a) and (b) we plot $E$ versus $w$ of Eq. (\ref{eq7}) for different 
cubic  ($p$) and quintic ($q$) nonlinearities. The energy minima of these plots correspond to a stable bullet
of negative energy. From Fig. {\ref{fig1}(a) we find that for $q=30$ such an energy minima 
exists for $p>20$. An accurate value of this limit can be obtained from  Eq. (\ref{eq7}): for 
$q=30$ this equation has solution for $p\ge p_{\mathrm{crit}}= 19.6$. 
Hence the NLS equation (\ref{eq1}) can have a stable light bullet solution 
for cubic nonlinearity $p$ greater than a critical value $p_{\mathrm{crit}}$. For $p< p_{\mathrm{crit}}$
the system is much too repulsive and is
not bound and escapes to infinity. However, this critical value $p_{\mathrm{crit}}$
of $p$ is a function of 
the quintic nonlinearity $q$. The $p_{\mathrm{crit}}-q$ correlation can be found from an 
attempt to solve Eq. (\ref{eq7}) numerically. The  $p_{\mathrm{crit}}-q$ correlation obtained in this fashion is plotted in Fig. \ref{fig1}(c). However, in addition to these stable bullets corresponding to a global minimum of energy with negetive energy 
values, one can also have metastable bullets corresponding to a local minimum of energy at positive energies. Such a situation is illustrated in Fig. \ref{fig1}(d) where we plot the variational $E-w$ curves for $p=10$ and different $q$ values. The bullet 
with $p=10$ and $q=1.3$ has a local minimum at a positive energy and is metastable in nature. In the following we will only consider the stable light bullets with negative energy.

Next we compare  in Fig. \ref{fig2}(a) the numerical and variational 
root-mean-square (rms) radius $r_{\mathrm{rms}}$ of a light bullet
versus cubic nonlinearity $p$ for different quintic nonlinearity $q=10,30$.
The variational result is given by:  $r_{\mathrm{rms}}= \sqrt{3/2}{w}$,
where $w$ is the equilibrium variational width.
In Fig. \ref{fig2}(b) we 
show the numerical and variational   energy $|E|$ of a light bullet 
versus $p$ for different $q$. {The numerical energy is calculated using Eq. (\ref{eq6a})
with the numerically obtained $\phi({\bf r},z)$.} 
 The energy of the light bullet is negative in 
all cases and its absolute value is plotted. 
For small $p$, the 
agreement between numerical and variational results is better. For
large $p$, the agreement between the 
two  is qualitative. For large nonlinearity $p$ in the NLS equation, the profile of the bullet deviates more from the Gaussian variational ansatz $-$ thus making the variational results more approximate.

\begin{figure}[!t]

\begin{center}
\includegraphics[width=\linewidth,clip]{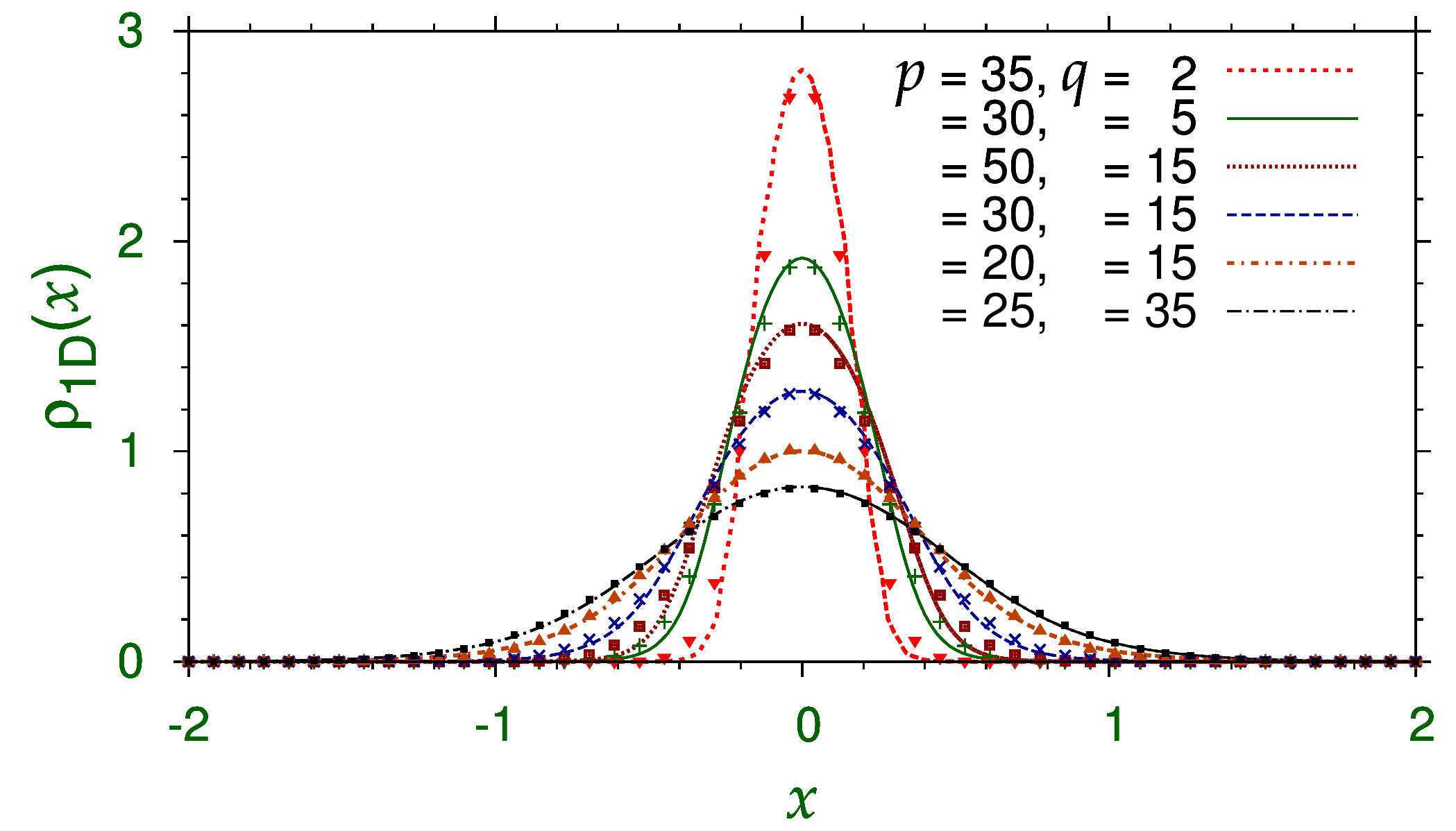}

\caption{(Color online) Numerical (line) and  variational (chain of symbols) reduced 1D density $\rho_{1D}(x)$ for different cubic nonlinearity $p$ and quintic nonlinearity $q$. 
}\label{fig3} 

\end{center}

\end{figure}

To study the density distribution of the light bullets we calculate the 
reduced 1D density defined by 
\begin{align}
\rho_{\mathrm{1D}}(x) = \int dt dy |\phi({\bf r})|^2.
\end{align}
In Fig. \ref{fig3} we plot this reduced 1D density as obtained from variational and numerical 
calculations for different cubic nonlinearity $p$ and quintic nonlinearity $q$.
For a fixed  defocusing nonlinearity
$q$ $(=15)$, the light bullet is more compact with the increase 
of focusing nonlinearity $p$ resulting in more attraction.  
For a fixed  focusing nonlinearity
$p$ $(=20)$, the light bullet is more compact with the decrease  
of defocusing nonlinearity $q$ resulting in less repulsion.

\begin{figure}[!t]

\begin{center}
\includegraphics[width=\linewidth,clip]{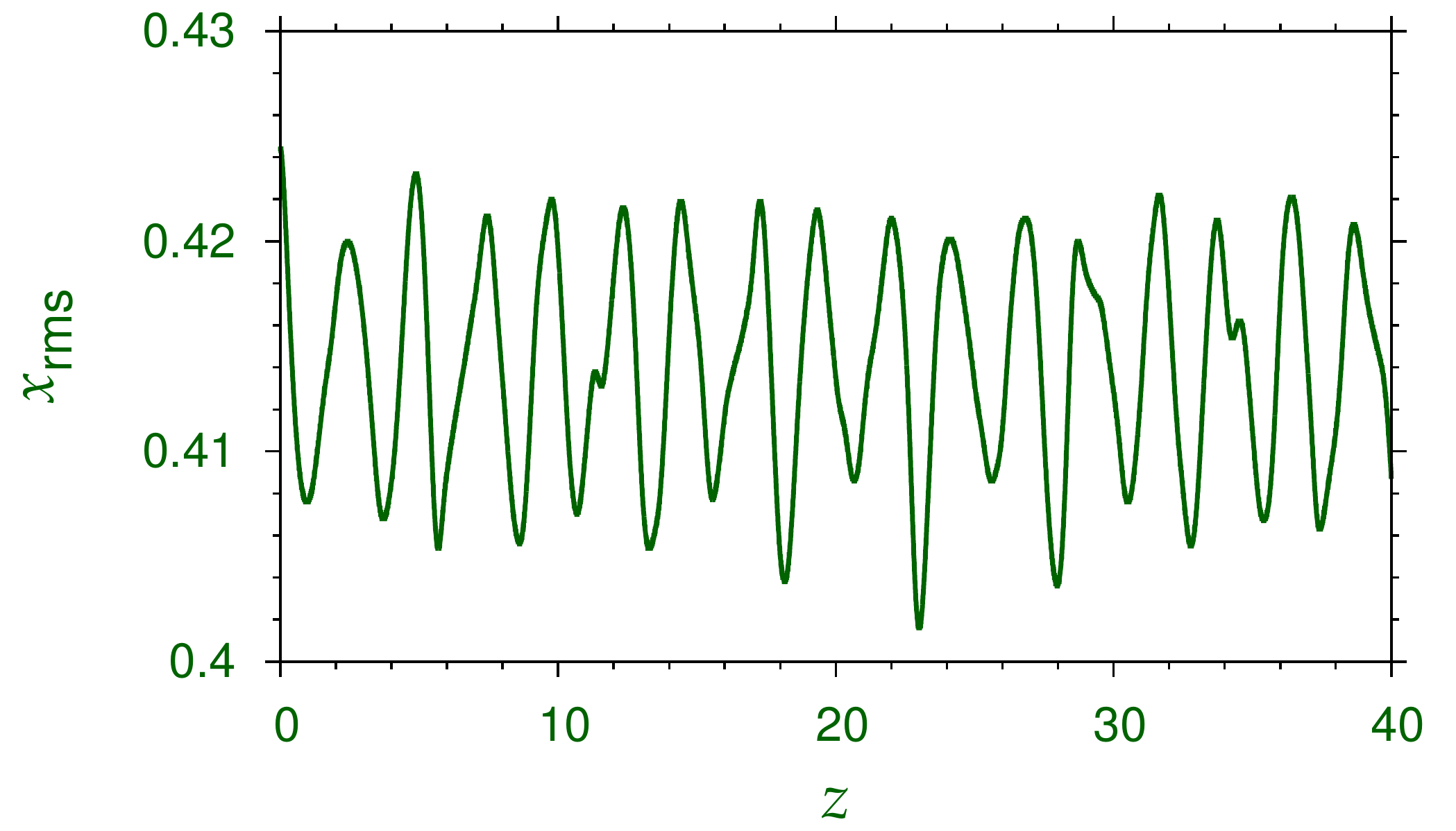}
 
\caption{(Color online)   Steady oscillation of the rms size $x_{\mathrm{rms}}$ 
during real-$z$ propagation of a light bullet with $p=20$ and $q=15$, when at $z=0$ 
the cubic nonlinearity $p$ is suddenly changed from 20 to 20.5.
}\label{fig4} 

\end{center}

\end{figure}

Now we present a numerical test of stability of a stable bullet 
under a small perturbation. For this purpose we consider the bullet shown in Fig. \ref{fig3}
with $p=20$ and $q=15$ as calculated by imaginary-$z$ propagation. Using the imaginary-$z$
profile as the initial state we perform numerical simulation by  real-$z$ propagation 
under a small perturbation introduced at $z=0$ by changing $p$ from 20 to 20.5. This sudden 
perturbation in the cubic nonlinearity  increases the attraction in the system and the light bullet starts a rapid breathing oscillation. In Fig. \ref{fig4} we show the steady oscillation in the rms 
$x$ size $x_{\mathrm{rms}}$ versus propagation distance $z$ during real-$z$ propagation. 
The steady continued oscillation of the bullet over a long distance of propagation establishes the stability of the bullet. The real-$z$ simulation was performed in full 3D
space without assuming spherical symmetry to guaranty the stability in full 3D Cartesian space.

 \begin{figure}[!t]

\begin{center}
\includegraphics[trim = 0mm 0mm 0mm 0mm, clip,width=.54\linewidth]{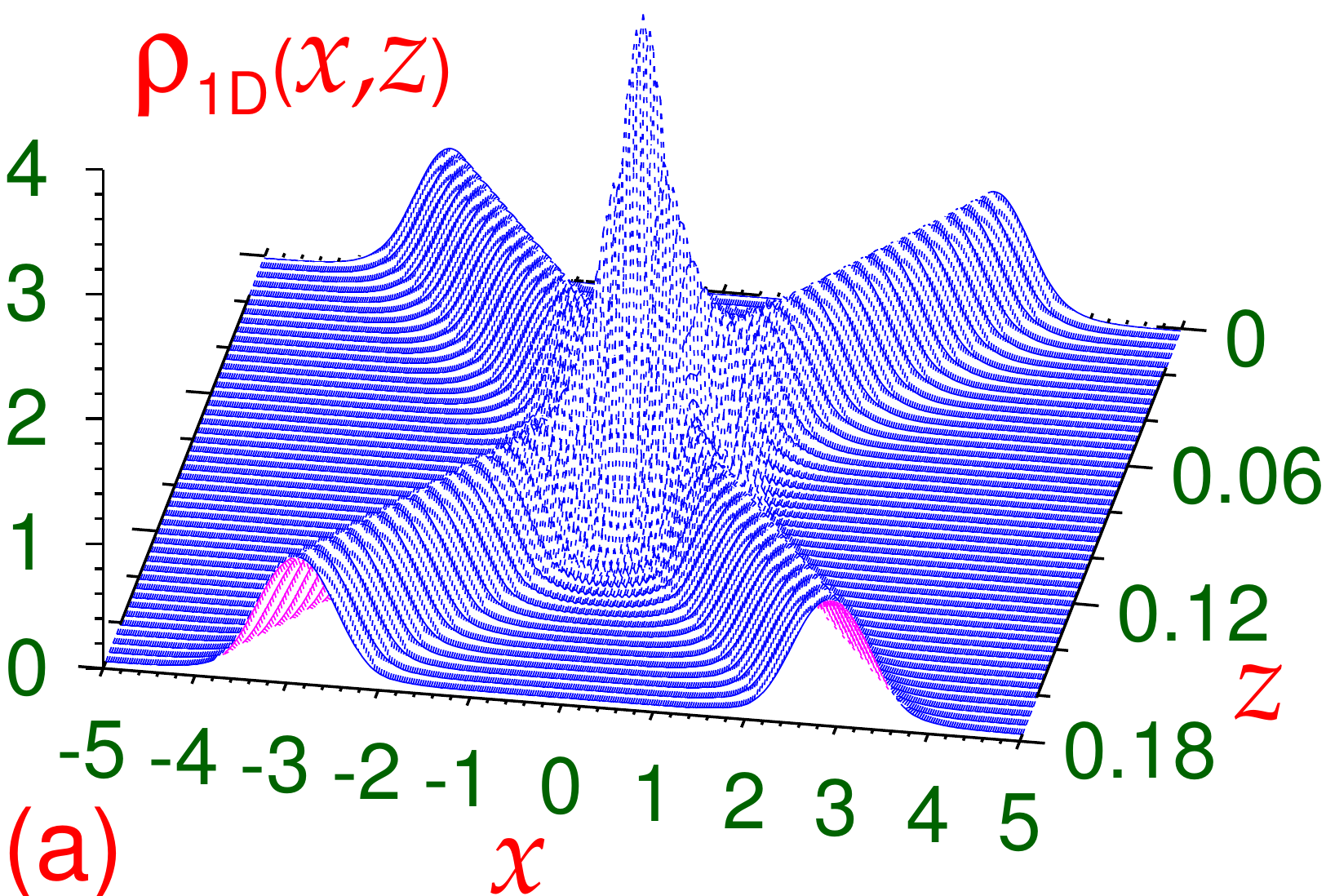}
 \includegraphics[trim = 0mm 2mm 0mm 2mm, clip,width=.44\linewidth]{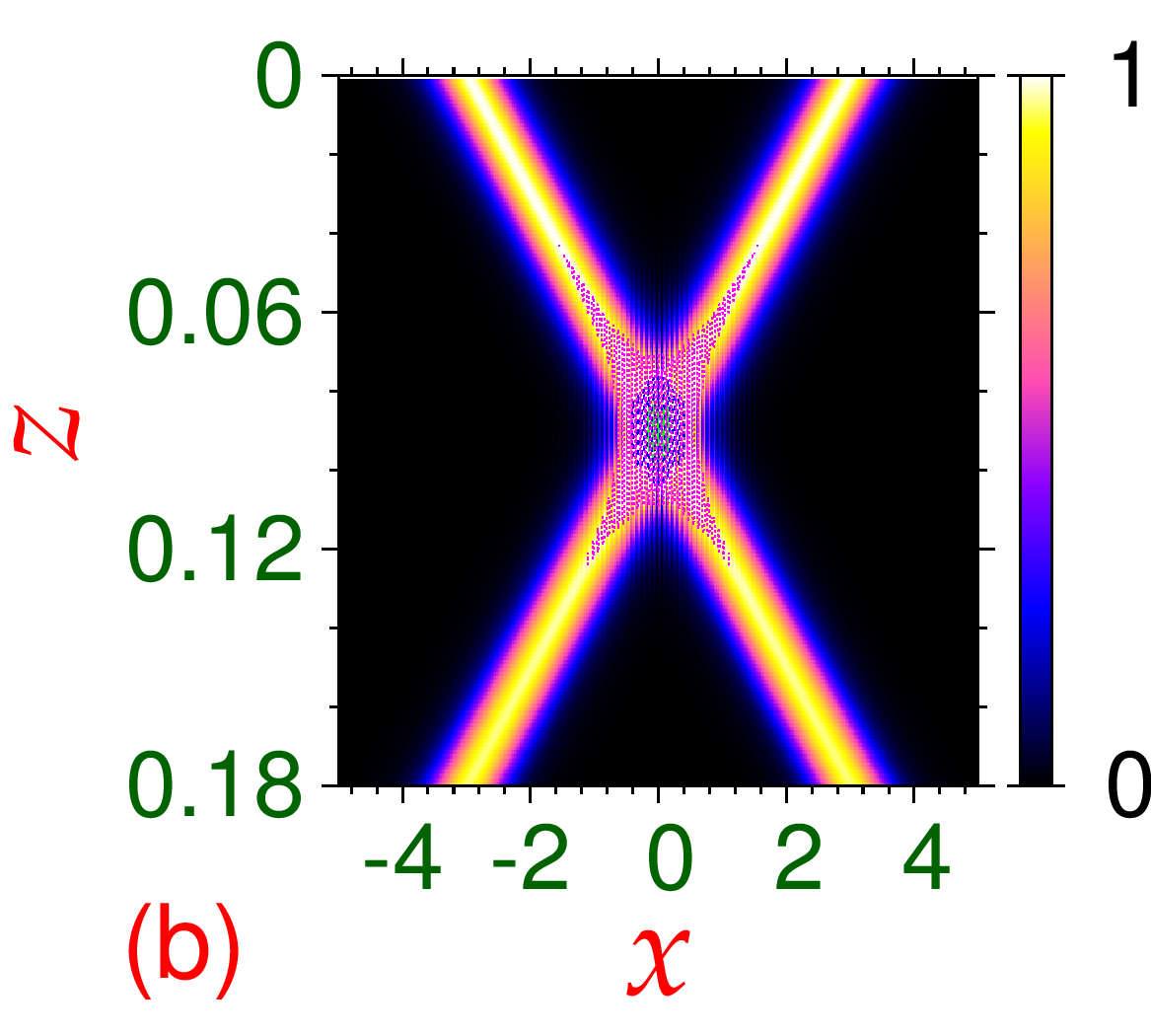} 
\includegraphics[trim = 0mm 0mm 0mm 0mm, clip,width=.54\linewidth]{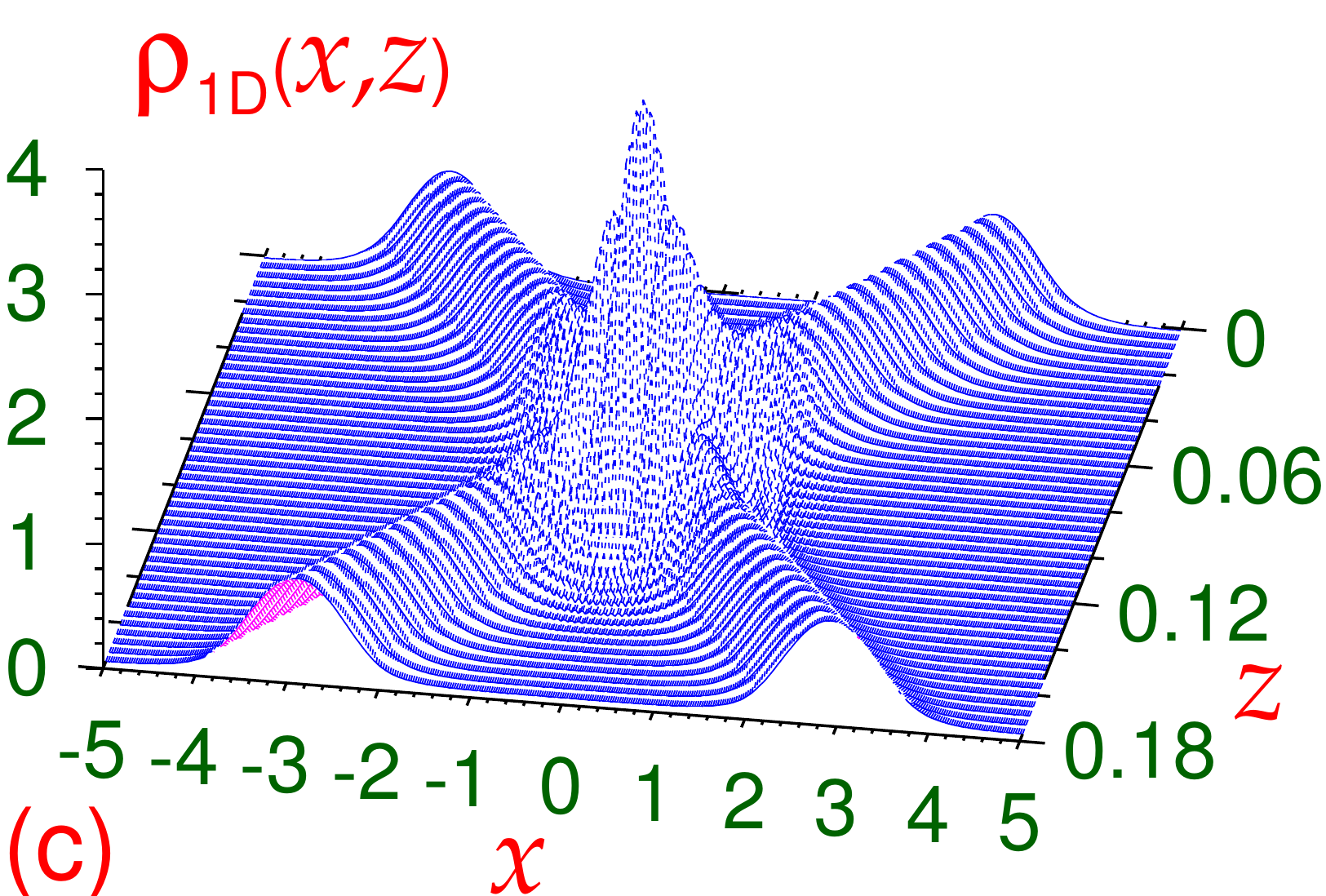}
 \includegraphics[trim = 0mm 0mm 0mm 2mm, clip,width=.44\linewidth]{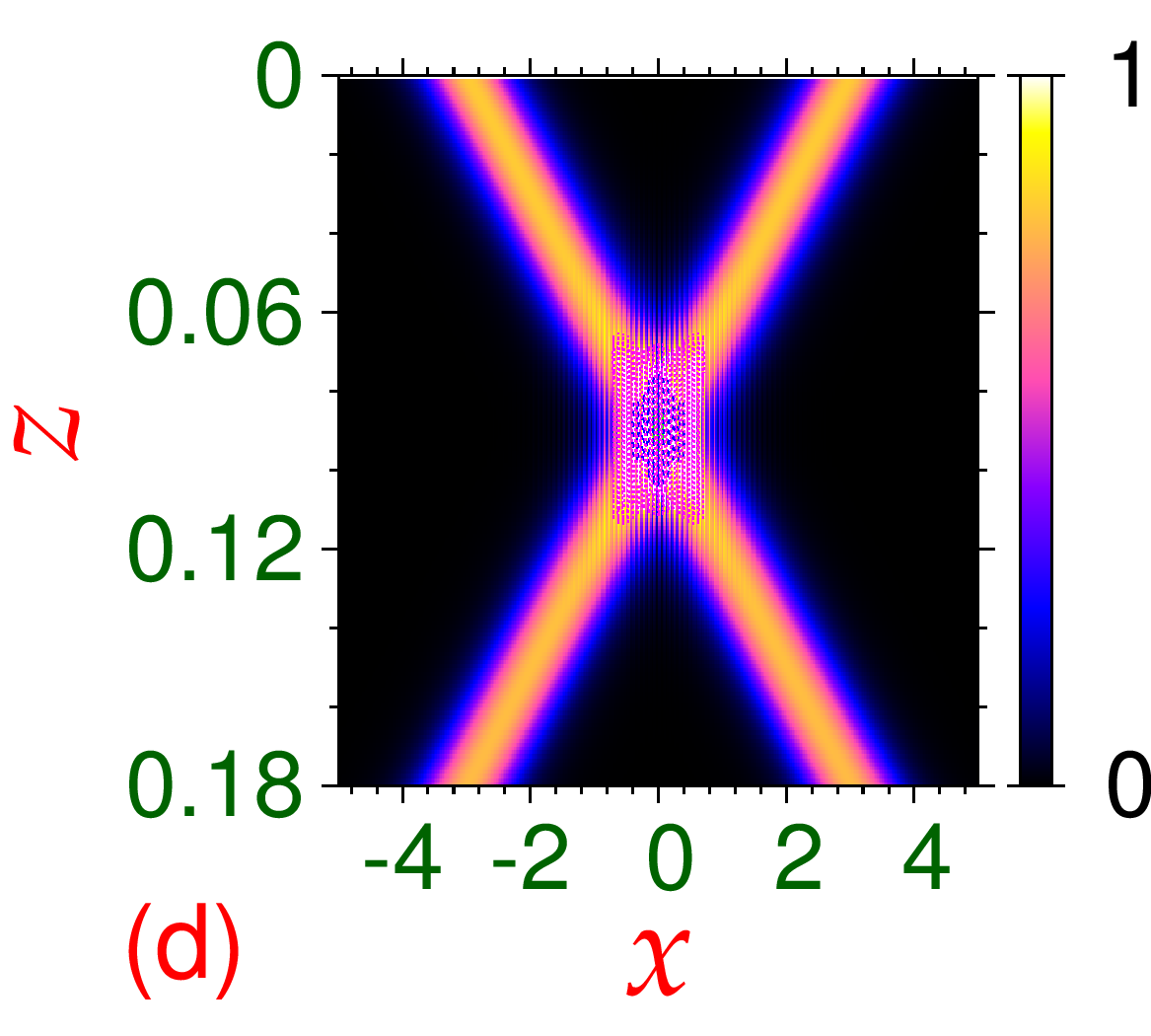}
\caption{ (Color online)
(a) The 1D density
$\rho_{1D}(x,z)$
 and (b) its contour plot during the 
collision of two light bullets     of Fig.  \ref{fig3} with $p=20,q=15$
 initially placed
at
$x=\pm 3$ , upon real-$z$ propagation. The initial wave functions are multiplied by
$\exp(\pm i40x)$ which sets them in motion with velocity of about $33$ each. The same for two light bullets with 
$p=25,q=35$
   of Fig.  \ref{fig3} are shown in (c) and (d). 
}\label{fig5} \end{center}

\end{figure}

 \begin{figure}[!t]

\begin{center}
\includegraphics[trim = 0mm 0mm 0mm 0mm, clip,width=.49\linewidth]{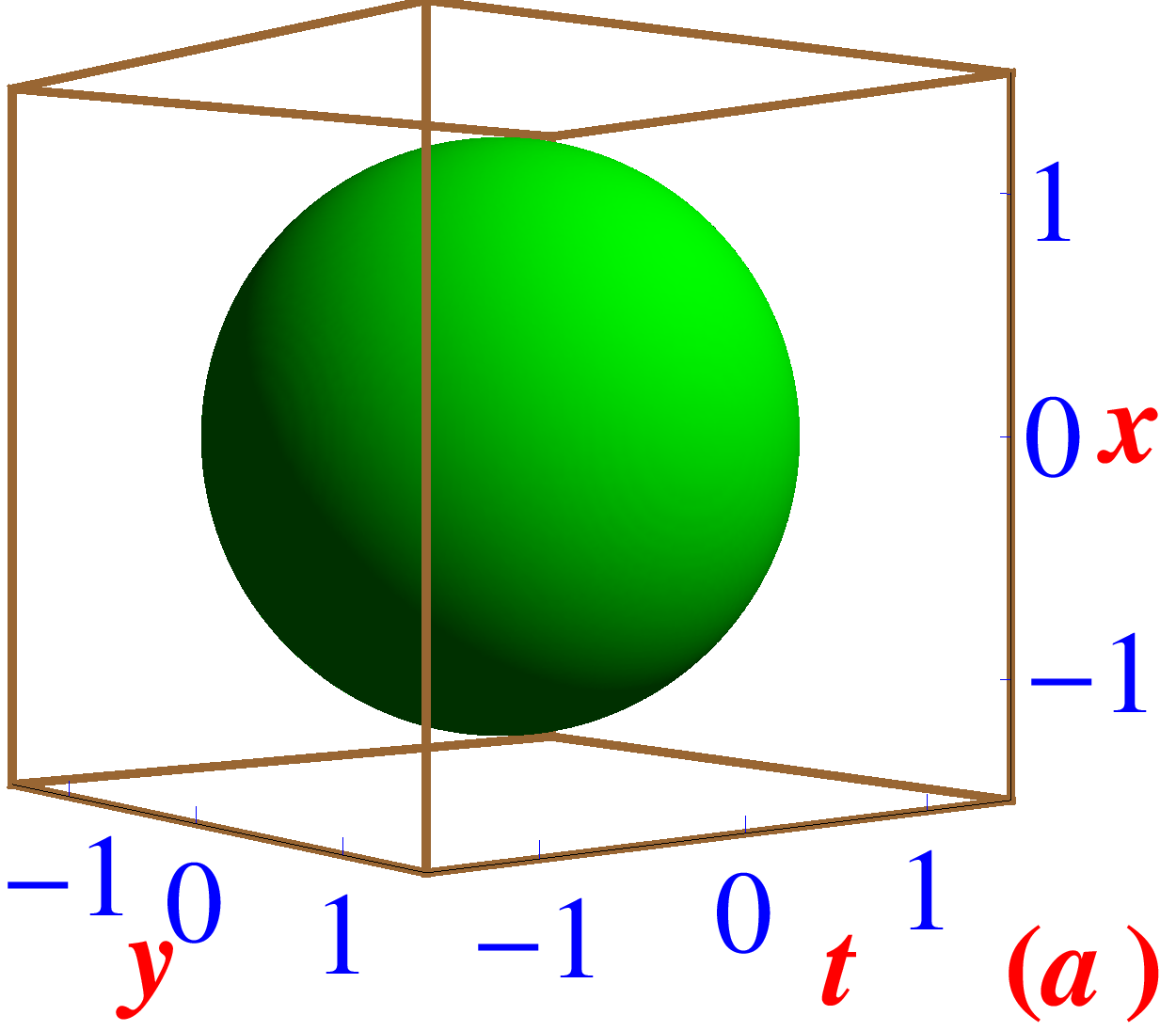}
 \includegraphics[trim = 0mm 0mm 0mm 0mm, clip,width=.49\linewidth]{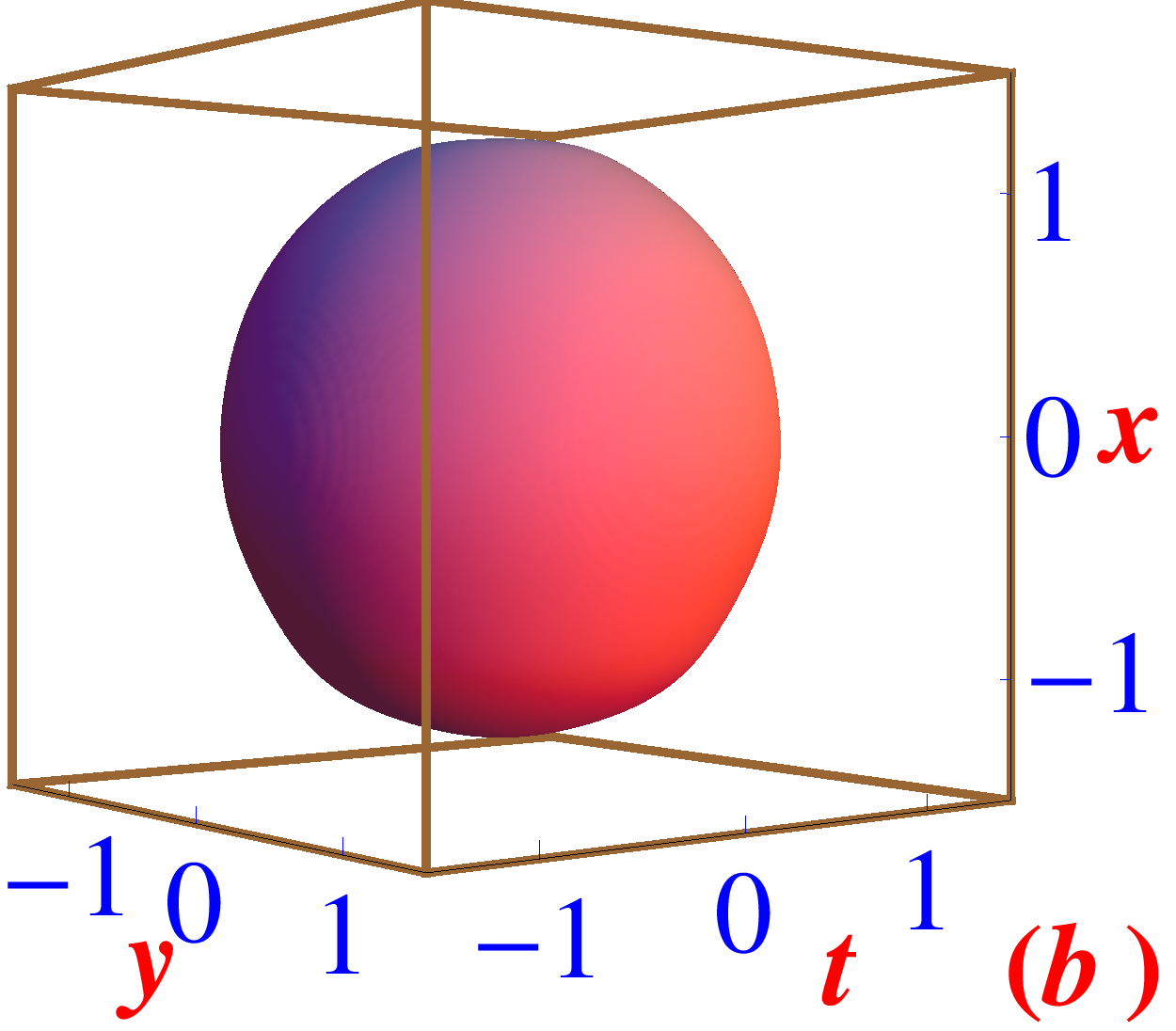} 
 \includegraphics[trim = 0mm 0mm 0mm 0mm, clip,width=.49\linewidth]{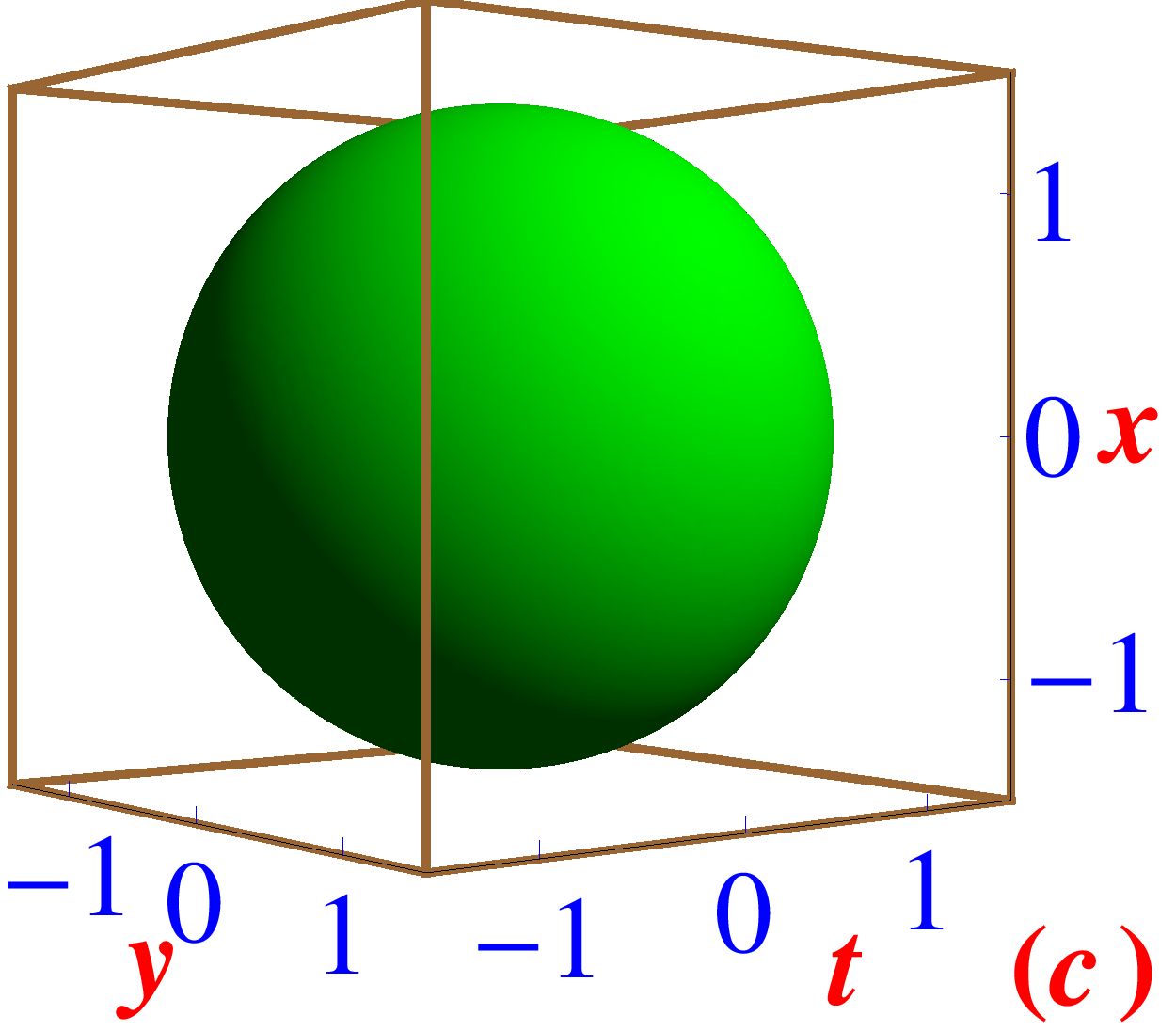}
 \includegraphics[trim = 0mm 0mm 0mm 0mm, clip,width=.49\linewidth]{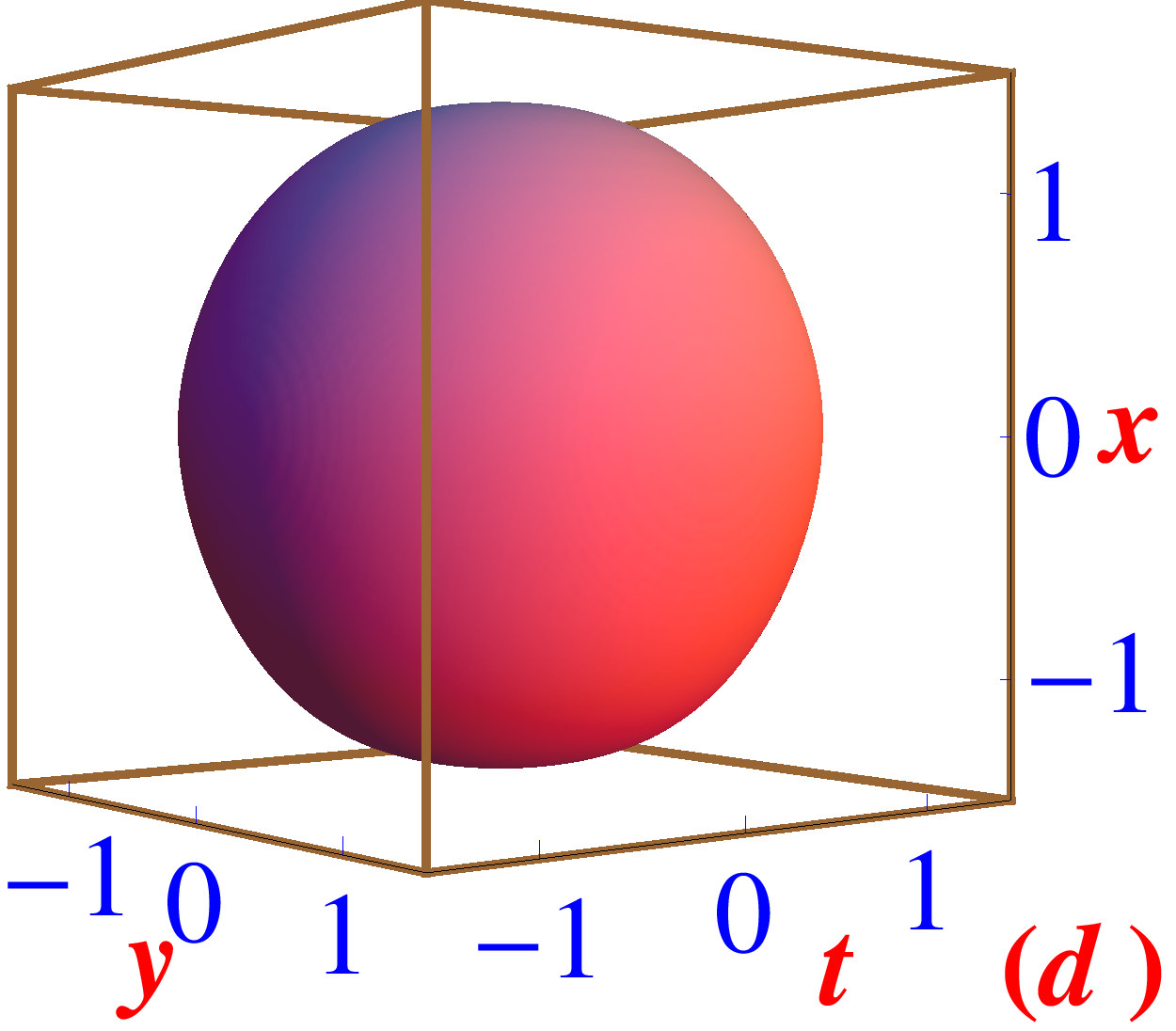}
\caption{ (Color online) 3D isodensity profile of the 
 (a) initial (at $z=0$) and (b) final (at $z=0.18$)  light bullets each with  $p=20, q=15$
and  (c) initial (at $z=0$) and (d) final (at $z=0.18$)  light bullets each with  $p=25, q=35$
 undergoing elastic collision 
 illustrated in Fig. \ref{fig5}(a) and (b). density on contour 0.01.
}\label{fig6} \end{center}

\end{figure}

The collision between two analytic  1D solitons is truly elastic \cite{rmp,book}     and such solitons pass through each other without deformation at any incident velocities.
The collision between  two 3D spatiotemporal light bullets is expected to be inelastic 
in general with loss of kinetic energy resulting in the  deformation of the bullets. 
Such a collision  can at best be quasi-elastic. To test the solitonic nature of the present light bullets,
 we study the frontal  head-on collision of two bullets.  
 The imaginary-$z$ profiles of the light bullets  shown in  Fig. \ref{fig3} with (i) $p=20, q=15$ and (ii) $p=25,q=35$  are  used as the initial function in the real-$z$ simulation of collision, with two identical bullets  placed at $x=\pm 3$ initially for $z=0$. 
To set the light bullets in motion along the $x$ axis in opposite directions the 
respective wave functions are multiplied by $\exp(\pm i 40x)$. To illustrate the dynamics upon real-$z$ simulation, we plot  the time evolution of  1D density $\rho_{1D}(x,z)\equiv \int dt\int dy |\phi({\bf r},z)|^2$ in Fig. \ref{fig5}(a) for the collision of two bullets with  $p=20,q=15$. The corresponding contour plot is presented in Fig. \ref{fig5}(b).    The same for the collision of two light bullets with  $p=25,q=35$ is presented in Figs.  \ref{fig5}(c) and (d).
The dimensionless velocity of a light bullet    is about 33 and the deviation from elastic collision is found to be small.
  Considering the three-dimensional nature of collision, the distortion in the bullet  profile is found to be negligible. 
To visualize the amount of inelasticity in the collision displayed in Figs. \ref{fig5}(a) and (b), we display in Figs. \ref{fig6}(a) and (b) the 3D isodensity profile of the light bullet before ($z=0$) and after ($z=0.18$) the collision shown in Figs. \ref{fig5}(a) and (b). 
 The same for the collision shown in Figs. \ref{fig5}(c) and (d) is illustrated in 
Figs. \ref{fig6}(c) and (d).  In general, the inelasticity in both 
collisions is small.

 \begin{figure}[!t]

\begin{center}
 \includegraphics[trim = 0mm 0mm 0mm 0mm, clip,width=.542\linewidth]{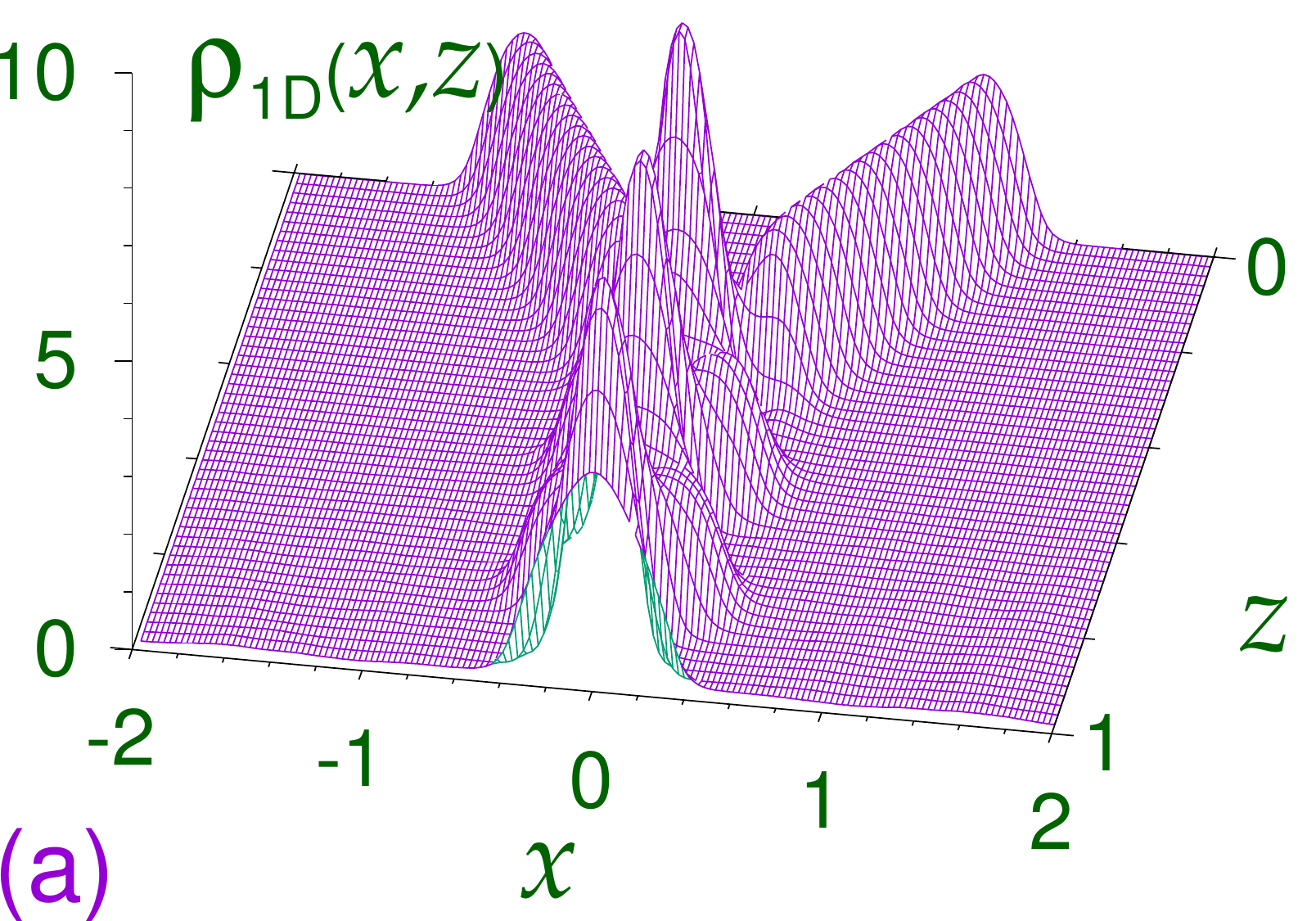}
  \includegraphics[trim = 0mm 0mm 0cm 0mm, clip,width=.442\linewidth]{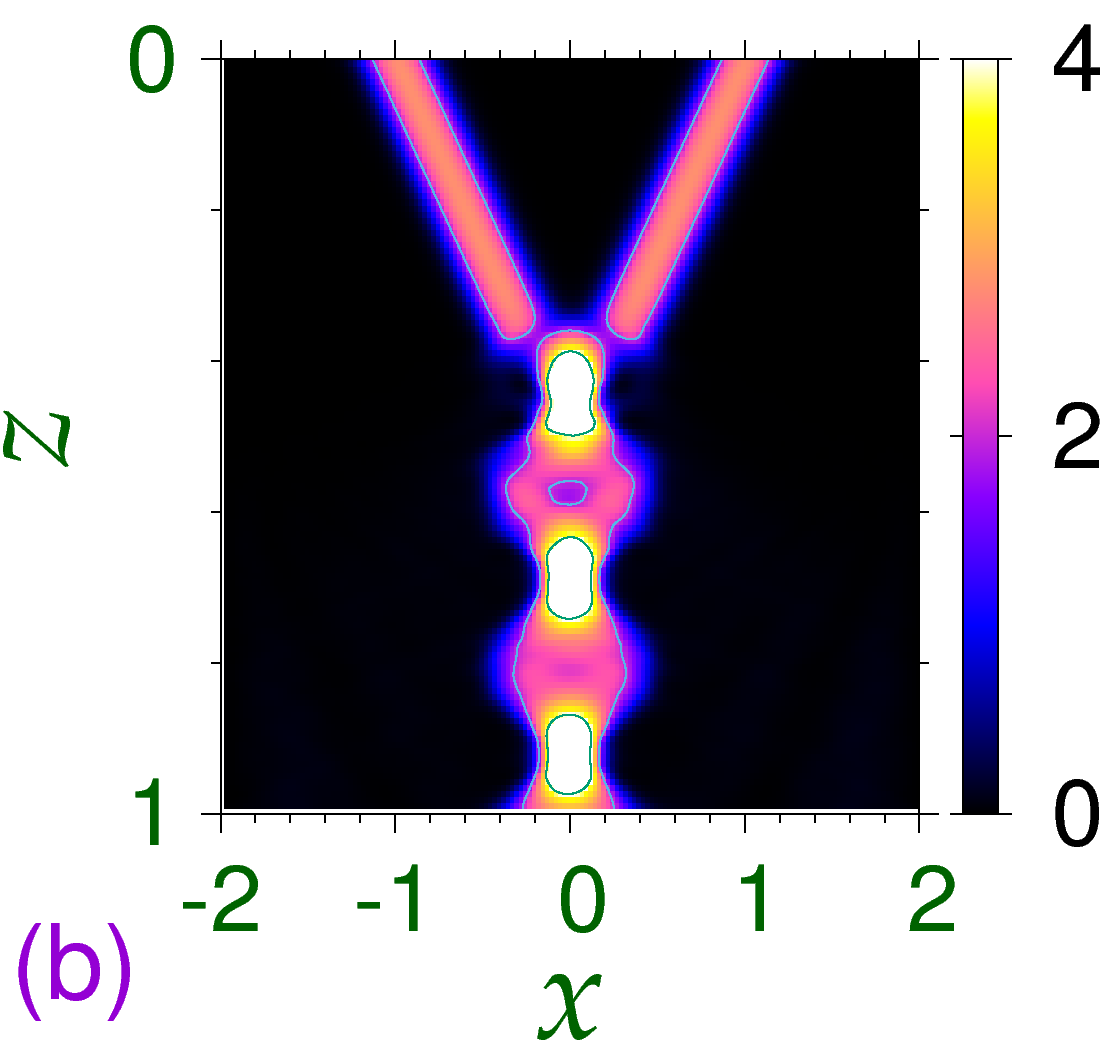}
 \includegraphics[trim = 0mm 0mm 3mm 0mm, clip,width=.542\linewidth]{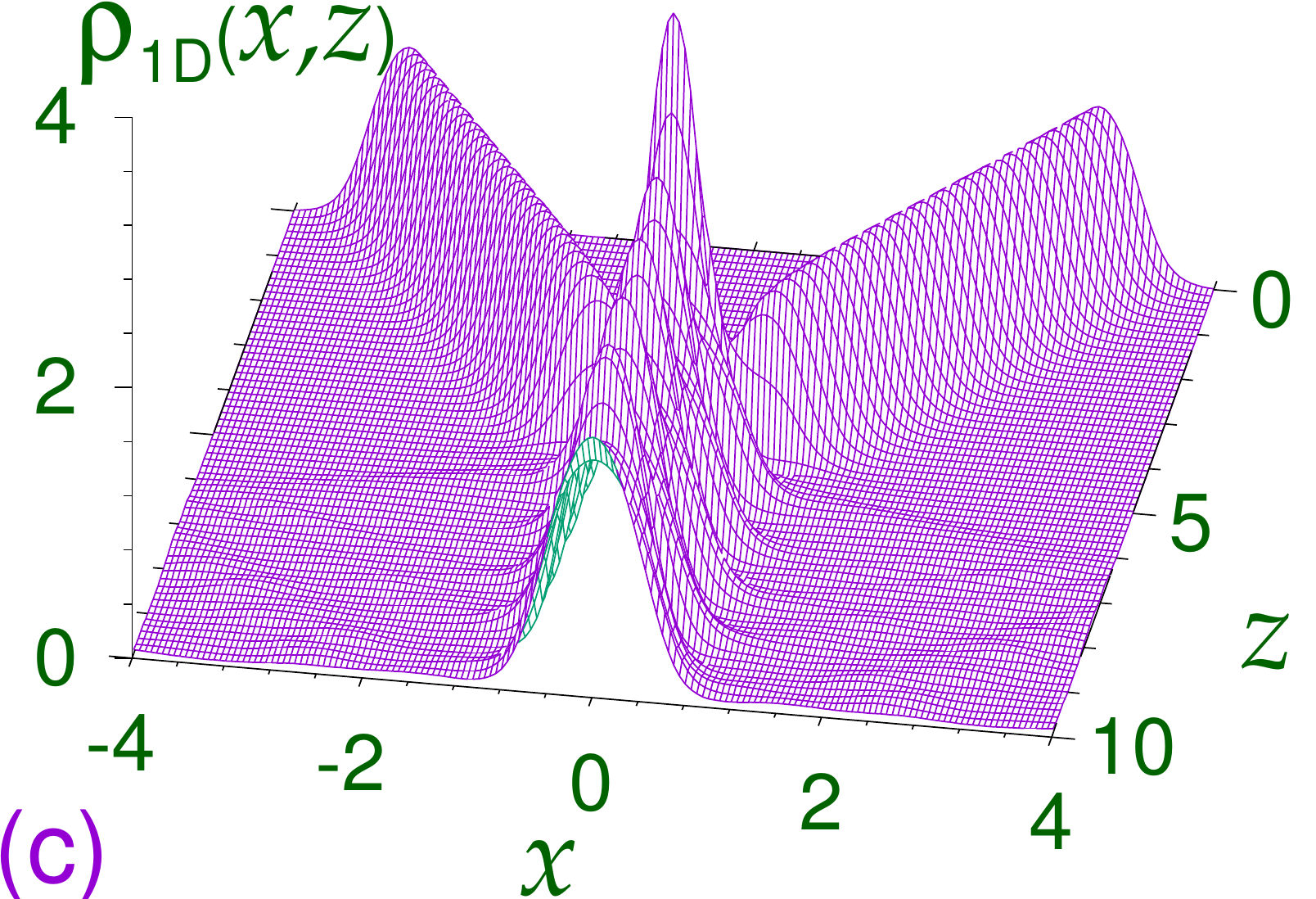}
  \includegraphics[trim = 0mm 0mm 1mm 0mm, clip,width=.442\linewidth]{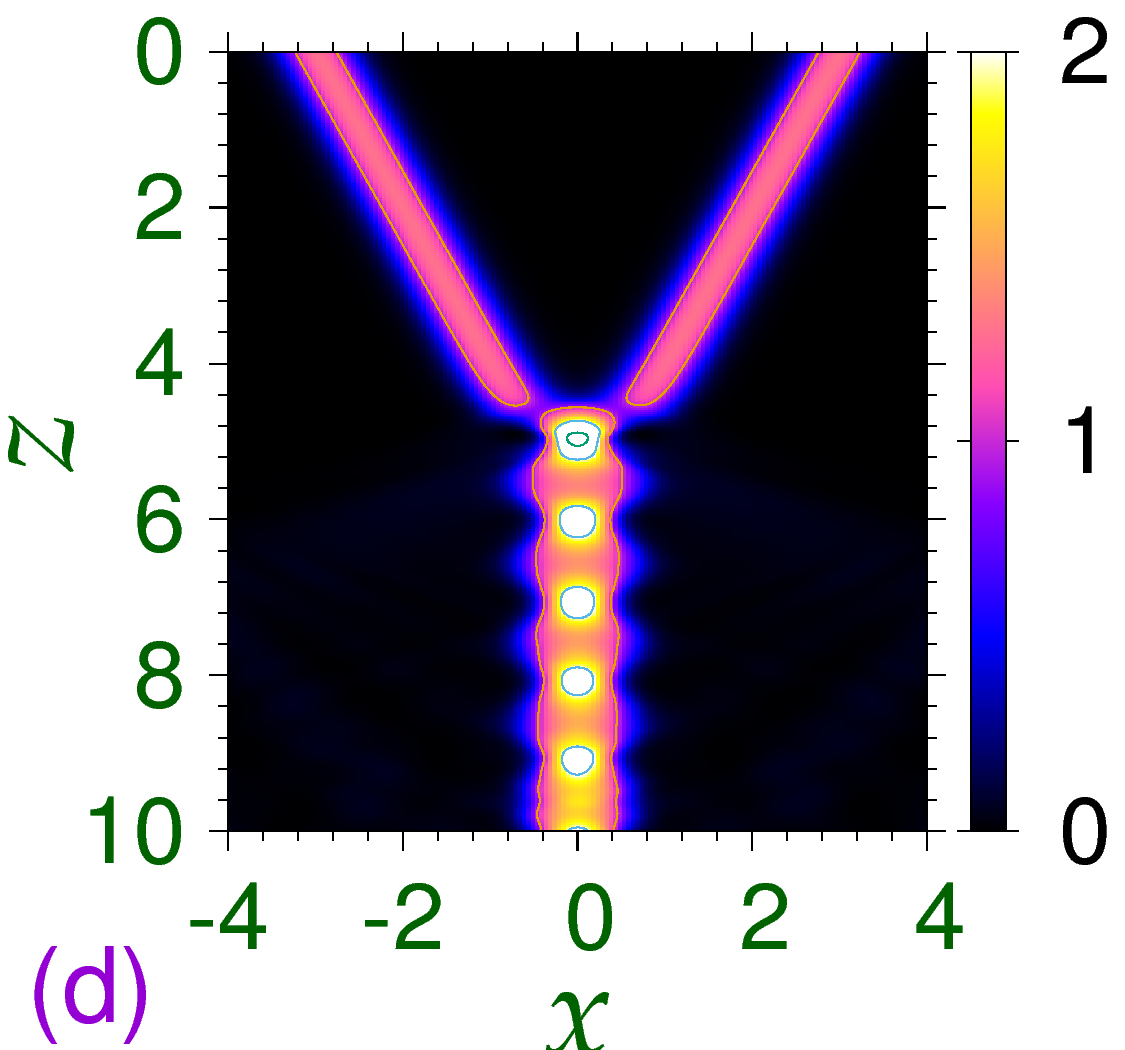}

\caption{ (Color online) 
(a) The 1D density $\rho_{1D}(x,z)$ and (b)
its contour plot during the collision of two light bullets of Fig.
\ref{fig3} with $p = 35, q = 2$ initially placed at $x =\pm  1$, upon 
real-$z$ propagation. The initial wave functions are multiplied by
$\exp(\pm i2x)$ to set them in motion with an initial dimensionless velocity of about 2. 
The same for the collision of two light
bullets with $p = 30, q = 15$ of Fig. \ref{fig3} for an initial velocity of about 
$0.5 $
are shown in (c) and (d), respectively.
}\label{fig7} \end{center}

\end{figure}

\begin{figure}[!b]

\begin{center}
\includegraphics[trim = 0mm 0mm 0mm 0mm, clip,width=.54\linewidth]{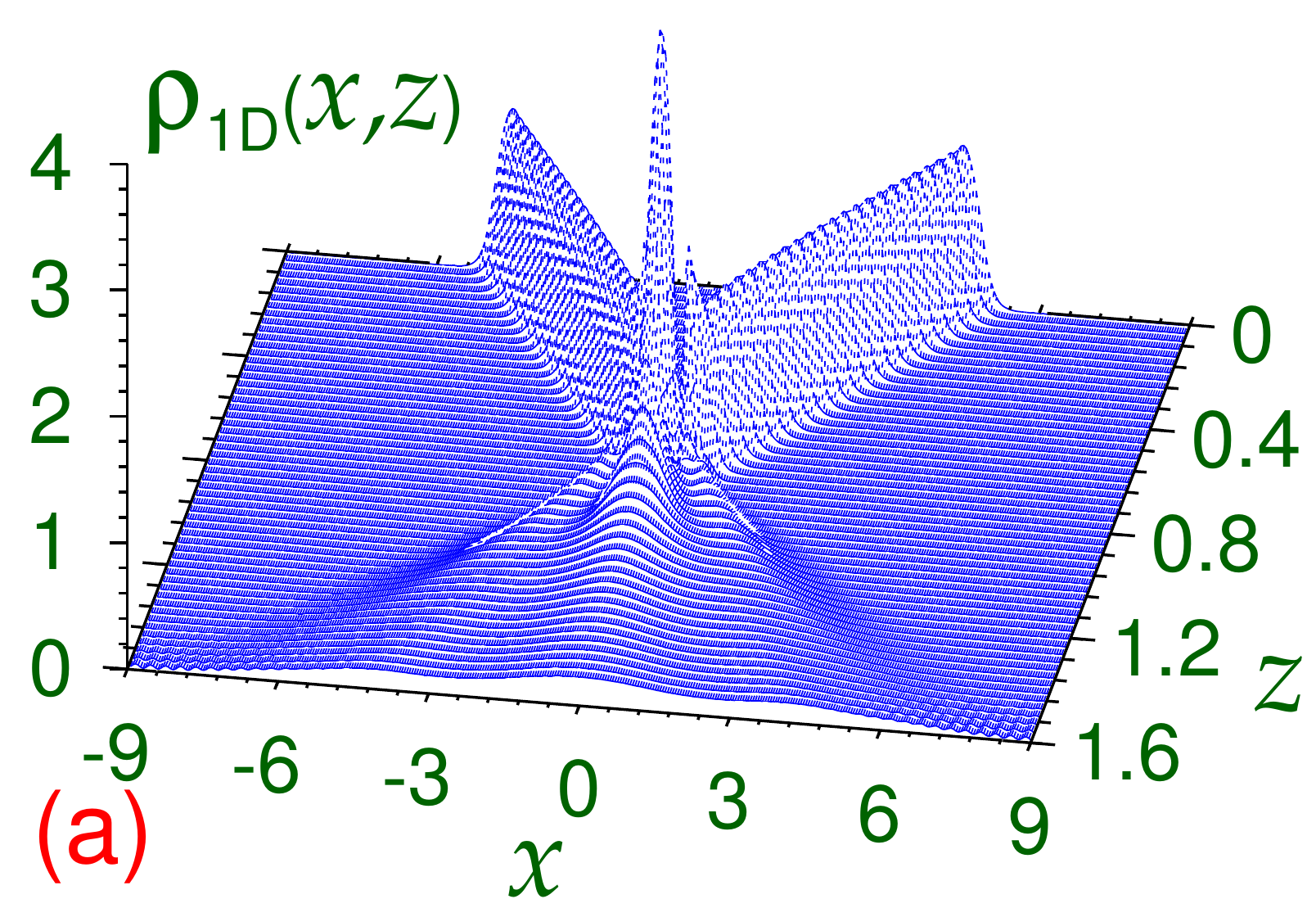}
 \includegraphics[trim = 0mm 2mm 0mm 2mm, clip,width=.44\linewidth]{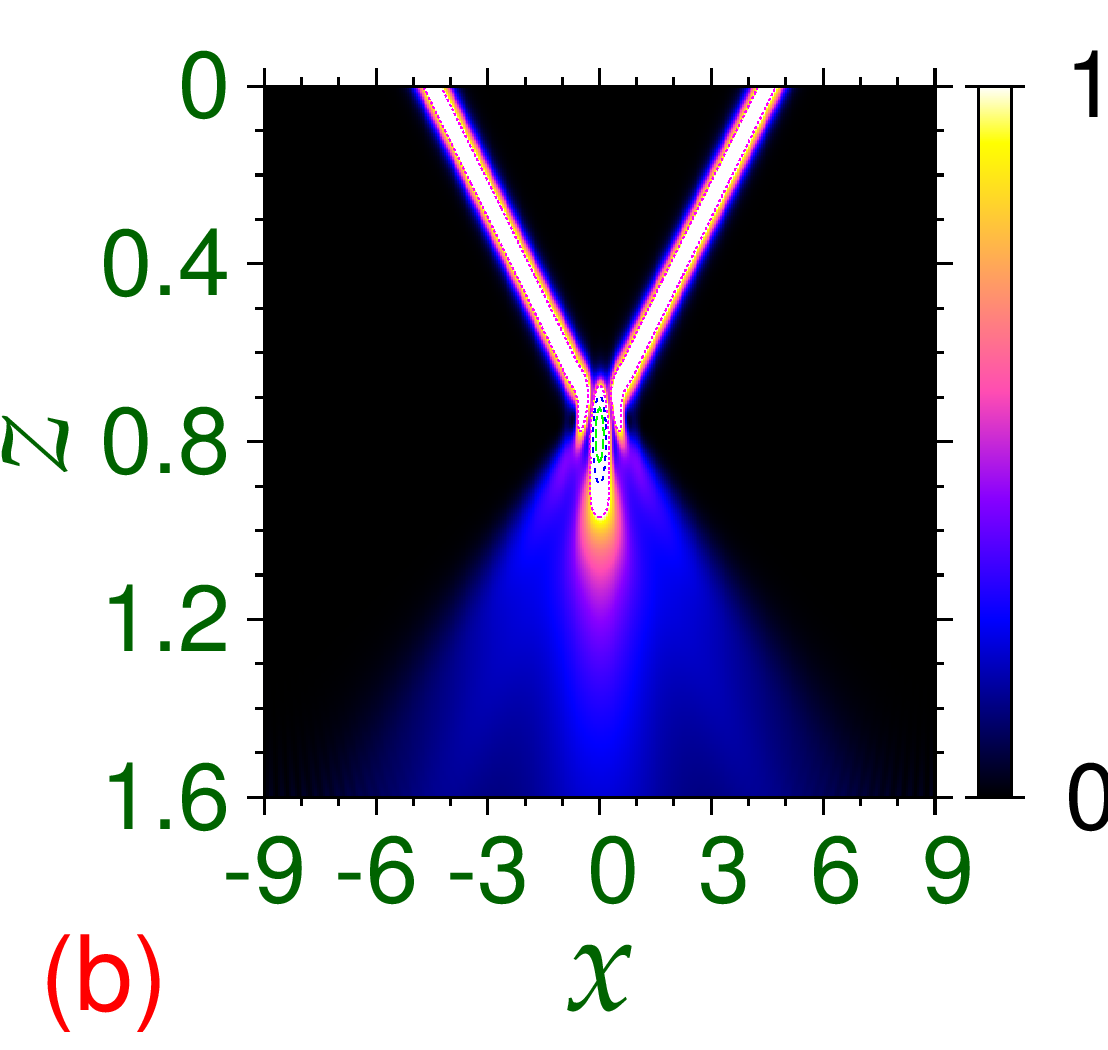}

\caption{ (Color online)
(a) The 1D density $\rho_{1D}(x,z)$ and (b)
its contour plot during the collision of two light bullets of Fig.
\ref{fig3} with $p = 30, q = 15$ initially placed at $x =\pm  4.5$, upon 
real-$z$ propagation. The initial wave functions are multiplied by
$\exp(\pm i6x)$ to set them in motion with an initial velocity of about $6$. 
}\label{fig8} \end{center}

\end{figure}

To study the inelastic collision   we consider two compact  bullets with 
$p=35,q=2$ and place them at $x=\pm 1$ and set them in motion with dimensionless velocity of about $2$ in opposite directions. This is achieved by multiplying the respective wave functions by $\exp(\pm i 2x$)
and perform real-$z$ simulation. The dynamics is illustrated by a plot of the time evolution of
1D density $\rho_{1D}(x,z)$ in Fig. \ref{fig7} (a) and the corresponding contour plot is shown in Fig. \ref{fig7} (b). The same for the collision of two light bullets with $p=30, q=15$
with an initial velocity of about $0.5$ are illustrated in Figs. \ref{fig7}(c) and (d),  
respectively.
In both cases    the two bullets come close to each other at $x=0$ 
coalesce to form a bullet molecule and never separate again. The combined bound 
system remain at rest at $x=0$ 
continuing small breathing oscillation because of a small amount of
 liberated kinetic energy which creates the bullet molecule in an excited state. 
{The observation of oscillating bullet molecules has been reported some
time ago in dissipative systems \cite{bm}.}

Hence at  sufficiently small incident velocities the collision of two light bullets lead 
to the formation of a  bullet molecule and at large velocities one has the quasi-elastic collision of two light bullets. At intermediate velocities a new phenomenon can take place. 
As the initial velocity is slowly increased from the domain of molecule formation, after collision a bullet molecule is formed in a highly excited state with a large amount of energy. In that case, because of the excess energy the bullet molecule   expands
and the localized bullets are destroyed. This is illustrated  by a plot of the time evolution of
1D density $\rho_{1D}(x,z)$ in Fig. \ref{fig8} (a) for the case of collision of two bullets 
with $p=30,q=15$ at an initial velocity of about $6$ 
 and the corresponding contour plot is shown in Fig. \ref{fig8} (b).

\section{Summary}

\label{IV}

Summarizing, we demonstrated the creation of a stable 3D spatiotemporal light bullet 
with cubic-quintic nonlinearity employing the Lagrange variational and full 3D numerical solution of the NLS equation. The statical properties of the bullet are studied by a 
variational approximation and a numerical imaginary-$z$ solution of the 3D NLS equation. 
 The cubic nonlinearity is  
taken as focusing Kerr type above a critical value, whereas the  quintic 
nonlinearity is defocusing.  
The dynamical properties are studied by a real-$z$ solution of the NLS equation. 
In the 3D spatiotemporal case, the
optical bullet  can move with a constant velocity.   At large velocities, the collision between the two 
spatiotemporal light bullets    is   quasi elastic with 
no visible deformation of the final bullets.  
At small velocities, the collision is inelastic with the formation of a bullet molecule
after collision. At medium velocities the bullets can be destroyed after collision.

\begin{acknowledgments} 
I thank F. K. Abdullaev for very valuable comments. 
Interesting discussion with Boris A. Malomed   is gratefully acknowledged. 
We thank the Funda\c c\~ao de Amparo 
\`a
Pesquisa do Estado de S\~ao Paulo (Brazil)
(Project:  2012/00451-0
  and  the
Conselho Nacional de Desenvolvimento   Cient\'ifico e Tecnol\'ogico (Brazil) (Project: 303280/2014-0) for 
support.
\end{acknowledgments}

%
\end{document}